\pdfoutput=1

\documentclass[%
 aip,
 jmp,
 amsmath,amssymb,
 reprint,%
 nofootinbib
]{revtex4-1}

\usepackage{hyperref}   
\begin{document}

\preprint{RCTP-1801}

\title{Quaternionic Gauge Transformations and Yang-Mills fields in Weyl Type Geometries}


\author{J.E. Rankin}
\email[]{physics@jrankin.users.panix.com}
\affiliation{San Miguel Road, Concord, CA}

\date{7 October 2019}

\begin{abstract}

This elementary discussion generalizes a Weyl geometry to allow quaternion valued gauge transformations and classical Yang-Mills geometric fields. This development will assume that the symmetric metric tensor is real in some gauge, and will develop the left and right handed approaches to quaternionic gauge transformations. Quaternionic gauge transformations are shown actually to require the shifting of some of Weyl's nonmetricity into torsion to define a properly transforming gauge field full curvature tensor, which is constructed as an asymmetric sum of left and right handed forms. Natural, gauge invariant, dimensionless variables are defined suitable for physics, and for use as a general formalism to describe these geometries, including General Relativity, in rather general circumstances. The geometry ``self measures'' these variables. Weyl's original action principle provides an example of an action rephrased in these gauge invariant variables, along with some unexpected possible insights on mechanics promoted by such a formulation of that action. Those include the torsion tensor and nonmetricity being constructed from mechanical energy-momentum. The Weyl form of action is then generalized to a quaternionic gauge field. The insights on mechanics now include spin 
$ 1 / 2 $ Dirac free fields. For physically reasonable choices of free parameters, the dimensionless Ricci tensor becomes nonnegligible in particle physics at distances much greater than the Planck length, along with limited general relativistic effects.
%

\end{abstract}

\maketitle


\renewcommand{\thefootnote}{\roman{footnote}}

\section{Introduction}

This work introduces a quaternion valued generalization of the Weyl geometry which can include General Relativity in a natural combination with quaternionic, classical Yang-Mills fields. Since quaternions have four components, and conventional 
$ SU(2) $ Yang-Mills fields operate with three\cite{guidry}, this is a more general Yang-Mills structure which actually contains the more standard 
$ SU(2) $ case as the subset with no real component. The extension of the Weyl geometry into quaternions will be seen to {\it require} the shifting of some nonmetricity into torsion in order to produce a nontrivial, correctly transforming, curvature tensor form associated with the Yang-Mills gauge field.

Exposition will be elementary such that anyone who can understand the original works of Weyl\cite{weyl.stm,adler.london} and Eddington\cite{eddington.mtr}, the basics of classical Yang-Mills fields\cite{guidry}, and elementary quaternions\cite{morse.feshbach,adler.qm} can follow this. Indeed, the core equations are (\ref{gauge.gamma.mixed.k}), (\ref{def.b.r}), (\ref{def.b.l}), (\ref{def.b.neat.k}), and (\ref{def.b.neat.gauge}), and those can probably be previewed and understood by many readers without reading the rest of this paper.

Additionally, when the scalar curvature is nonzero, the structure itself defines intrinsic, simple, gauge invariant, dimensionless variables that make it possible to carry through an exposition relatively simply in the quaternions. The structure literally measures itself via these variables.\footnote{This use of the word ``measure'' does {\it not} include the eventual reduction of a superposition of states into an eigenstate. It refers only to the fact that the geometry intrinsically defines the overall quantities to be eventually measured experimentally in the laboratory.} Without them, the division into right and left handed expressions, and the presence of noncommuting, quaternionic quantities will produce cumbersome expressions, or quantities that are difficult to interpret. These gauge invariant variables will be seen always to obey a necessary kinematic constraint imposed by the geometry. Substantial similarities will be found between that constraint equation, and well known equations from classical and quantum particle mechanics, including those of spin 
$ 1 / 2 $ Dirac particles.

Specific action principles for the structure will be examined, first for Weyl's originally proposed unified field dynamics restricted to the real numbers, and then for a fully quaternionic, Yang-Mills type gauge field combined with standard General Relativity, with optional additional terms in the action suggested by Weyl's original action choice. Those actions will produce the examples of similarities between the constraint and forms familiar from particle mechanics.

In all expressions, unless noted otherwise, the partial derivative with respect to coordinate 
$ x^{\mu} $ is simply denoted by 
``$ {}_{, \mu} $'', and 
$ x^{\mu} $ itself will be considered to be rendered dimensionless through the introduction of a universal scale factor 
$ b_{0} $. That scale factor relates dimensionless Lorentzian coordinates 
$ x^{\mu} $ to lab unit Lorentzian coordinates 
$ x^{\mu}_{LAB} $ via 
$ x^{\mu} = \sqrt{ b_{0} } \, x^{\mu}_{LAB} $. The constant 
$ b_{0} $ is an inverse length squared, one that is not assigned a specific value initially, but is assumed to be definite. Ideally, an obvious value would eventually emerge through comparison of equations to physics.

\section{Quaternionic Gauges and Curvatures in Weyl-like Geometries}
\label{extended}

In order to make sense of Weyl-like geometries allowing quaternionic gauge transformations, one foundation rule is necessary. There must exist a gauge or gauges in which the metric tensor is real (with signature 
$ ( + - - - ) $). Such a metric is considered to be a given. That simplifies the model enough to proceed fairly easily.

\subsection{Quaternions}

In this presentation, the basic quaternions will be taken as abstract mathematical objects similar to the number 
$ 1 $ and the imaginary unit $ \imath $ 
in complex numbers. Specifically, they are taken to be the four quantities 
$ Q_{\mu} $, where the subscript does {\it not} imply the quantities are a four vector, and where
\begin{equation}
Q_{0} = \sigma_{0}
\label{def.q0}
\end{equation}
and
\begin{equation}
Q_{k} = - \imath \sigma_{k}
\label{def.qk}
\end{equation}
for $ k = 1,2,3 $, where the $ \sigma_{k} $ 
are the standard Pauli spin matrices\cite{schiff}, and 
$ \sigma_{0} $ is the unit matrix. The basic properties of quaternions are reviewed in many references, such as Adler\cite{adler.qm}, and Morse and Feshbach\cite{morse.feshbach}, and there are many representations of them which may differ from those of equations (\ref{def.q0}) and (\ref{def.qk}), yet which are algebraically isomorphic to those quantities. Such possibly isomorphic representations will be denoted here by 
$ Q^{\prime}_{\mu} $ if needed, and they could be given particular coordinate transformation properties for convenience, unlike the 
$ Q_{\mu} $, which are mathematical invariants.

The $ Q_{k} $ have an obvious vector form
\begin{equation}
\vec Q = \sum_{ k = 1 }^{3} \hat e_{k} Q_{k} 
\label{def.qvec}
\end{equation}
where the $ \hat e_{k} $ 
are the Cartesian unit vectors. A completely analogous form exists for the 
$ Q^{\prime}_{k} $,
\begin{equation}
\vec Q^{\prime} = \sum_{ k = 1 }^{3} \hat e^{\prime}_{k} Q^{\prime}_{k} 
\label{def.qvecprime}
\end{equation}
although now the 
$ \hat e^{\prime}_{k} $ may be unit vectors in one of the curvilinear coordinate systems in common use\cite{morse.feshbach}. A general quaternion 
$ A $ is given by 
$ A = Q_{0} A_{0} + \vec Q {\bf \cdot} \vec A $ where 
$ A_{0} $ is called the real part, and the vector 
$ \vec A $ is called the imaginary part. These quaternion forms are called real quaternions because 
$ A_{0} $ and 
$ \vec A $ are themselves real quantities.

\subsection{Basics of the Geometry}

The basic Weyl-like geometry is understood to be based on a symmetric metric tensor and a Weyl four vector combined into a more general affine connection\cite{weyl.stm,eddington.mtr}. The reason for the term ``Weyl-like'' here in place of just ``Weyl'' will become apparent shortly, when it will be seen that the connection must include some torsion through an unbalancing of the original Weyl connection in order for the overall geometry to be nontrivial and reasonably neat in the quaternions.

In the gauge in which the metric is real, 
$ g_{\mu \nu} $ will be denoted by 
$ \tilde g_{\mu \nu} $. Tensor indices are lowered and raised using 
$ g_{\mu \nu} $, and its inverse.

Since $ g_{\mu \nu} $ is real in some gauge, it is always possible to write
\begin{equation}
g_{\mu \nu} = \tilde g_{\mu \nu} \gamma 
\label{g.to.greal}
\end{equation}
where $ \gamma $ is some quaternionic scalar. This allows the various operations with 
$ g_{\mu \nu} $ to be set up through easy correspondence with operations on real forms. For example, 
$ g^{\mu \nu} $ is easily calculated through the standard requirement that
\begin{equation}
g^{\mu \alpha} g_{\alpha \nu} = \delta^{\mu}_{\nu} 
\label{g.ginv}
\end{equation}
where $ \delta^{\mu}_{\nu} $ is the Kronecker delta. In fact,
\begin{equation}
g^{\mu \nu} = \gamma^{-1} \tilde g^{\mu \nu} 
\label{ginv.to.ginvreal}
\end{equation}
Furthermore, equations (\ref{g.to.greal}) and (\ref{ginv.to.ginvreal}) give
{\samepage 
\begin{eqnarray}
g^{\alpha \tau} \! g_{\mu \nu} & = & 
\gamma^{-1} \tilde{g}^{\alpha \tau} \tilde{g}_{\mu \nu} \gamma
\nonumber \\
 & = & 
\tilde{g}^{\alpha \tau} \tilde{g}_{\mu \nu} 
\nonumber \\
 & = & 
\tilde{g}_{\mu \nu} \tilde{g}^{\alpha \tau} 
\nonumber \\
 & = & 
\tilde{g}_{\mu \nu} \gamma \gamma^{-1} \tilde{g}^{\alpha \tau} 
\nonumber \\
 & = & 
g_{\mu \nu} g^{\alpha \tau} 
\label{gauge.inv.gpair}
\end{eqnarray}}%
That metric combination is always real, gauge invariant, and commutes {\it as a unit} with everything.

Since gauge transformations are to be quaternionic, they can generally be applied to 
$ g_{\mu \nu} $ from either the left or right, as denoted by
\begin{equation}
\bar g_{\mu \nu} = \lambda g_{\mu \nu} \rho 
\label{gbar.lr}
\end{equation}
where $ \lambda $ is the left gauge transformation, and 
$ \rho $ is the right. However, more restricted cases where one of these two multipliers is always taken to be 
$ 1 $ will be most useful here. Since equation (\ref{gbar.lr}) can always be written as
\begin{equation}
\bar g_{\mu \nu} = \lambda \tilde g_{\mu \nu} \gamma \rho 
\label{gbar.lr.gamma}
\end{equation}
according to equation (\ref{g.to.greal}), and since 
$ \tilde g_{\mu \nu} $ commutes with every scalar multiplier, restricting either 
$ \lambda $ or $ \rho $ to be 
$ 1 $ may not be a serious restriction in practice.

\subsection{Left and Right Covariant Derivatives / Christoffel Symbols}

Since $ g_{\mu \nu} $ can now have a limited quaternionic nature, the expression
\begin{equation}
g_{\mu \nu ; \gamma} = g_{\mu \nu , \gamma} 
- g_{\alpha \nu} \left \{ \left. {}^{\: \alpha}_{\mu \gamma } \right ] \right. 
- g_{\mu \alpha} \left \{ \left. {}^{\: \alpha}_{\nu \gamma } \right ] \right. 
= 0 
\label{def.right.cov.metric}
\end{equation}
is not necessarily the same as
\begin{equation}
g_{\mu \nu ; \gamma} = g_{\mu \nu , \gamma} 
- \left [ \left. {}^{\: \alpha}_{\mu \gamma } \right \} \right. g_{\alpha \nu} 
- \left [ \left. {}^{\: \alpha}_{\nu \gamma } \right \} \right. g_{\mu \alpha} 
= 0 
\label{def.left.cov.metric}
\end{equation}
Thus, equation (\ref{def.right.cov.metric}) defines the covariant derivative of 
$ g_{\mu \nu} $ with respect to the ``right handed Christoffel symbol'' 
$ \left \{ \left. ^{\: \alpha}_{\mu \nu} \right ] \right. $, while equation (\ref{def.left.cov.metric}) defines the covariant derivative of 
$ g_{\mu \nu} $ with respect to the ``left handed Christoffel symbol'' 
$ \left [ \left. ^{\: \alpha}_{\mu \nu} \right \} \right. $. Both equations actually define the associated Christoffel symbols, giving
\begin{equation}
\left \{ \left. ^{\: \alpha}_{\mu \nu} \right ] \right. = 
\frac{1}{2} \; g^{\alpha \tau} ( g_{\mu \tau , \nu} + 
g_{\nu \tau , \mu} - g_{\mu \nu , \tau} ) 
\label{def.right.chrst}
\end{equation}
and
\begin{equation}
\left [ \left. ^{\: \alpha}_{\mu \nu} \right \} \right. = 
\frac{1}{2} \: ( g_{\mu \tau , \nu} + 
g_{\nu \tau , \mu} - g_{\mu \nu , \tau} ) \, 
g^{\alpha \tau} 
\label{def.left.chrst}
\end{equation}
respectively.

Furthermore, equation (\ref{g.ginv}) and 
$ \delta^{\mu}_{\nu , \tau} = 0 $ give
\begin{equation}
g^{\beta \mu}_{\; \; \; \; , \tau} = 
- g^{\beta \nu} g_{\nu \alpha , \tau} \, g^{\alpha \mu} 
\label{inv.met.der}
\end{equation}
This and equations (\ref{gauge.inv.gpair}), (\ref{def.right.cov.metric}), and (\ref{def.left.cov.metric}) then give
\begin{equation}
g^{\beta \xi}_{\; \; \; \; ; \tau} \equiv 
g^{\beta \xi}_{\; \; \; \; , \tau} 
+ \left \{ \left. {}^{\: \beta }_{\alpha \tau } \right ] 
\right. g^{\alpha \xi} + \left \{ \left. 
{}^{\: \xi }_{\alpha \tau } \right ] \right. g^{\beta \alpha} 
= 0 
\label{def.right.cov.imetric}
\end{equation}
and
\begin{equation}
g^{\beta \xi}_{\; \; \; \; ; \tau} \equiv 
g^{\beta \xi}_{\; \; \; \; , \tau} + g^{\alpha \xi} 
\left [ \left. {}^{\: \beta }_{\alpha \tau } \right \} 
\right. + g^{\beta \alpha} \left [ \left. 
{}^{\: \xi }_{\alpha \tau } \right \} \right. 
= 0 
\label{def.left.cov.imetric}
\end{equation}
This indicates that the contravariant metric's indices interact with their associated Christoffel symbols on the opposite side from the covariant metric's indices in the definition of the covariant derivative of the metric. That same positioning convention for covariant and contravariant indices and their associated connection is now adopted here for all covariant derivatives of {\it any} tensor quantity as well, and also for more general affine derivatives, where the affine derivative simply uses the full affine connections which also include the Weyl four vector\cite{weyl.stm}, 
$ {}_{R} \Gamma^{\alpha}_{\mu \nu} $ and 
$ {}_{L} \Gamma^{\alpha}_{\mu \nu} $ in place of the Christoffel symbols 
$ \left \{ \left. ^{\: \alpha}_{\mu \nu} \right ] \right. $ and 
$ \left [ \left. ^{\: \alpha}_{\mu \nu} \right \} \right. $. The more general affine connections are prefixed with the lowered ``R'' and ``L'' to maintain different right and left handed forms on the more general level, just as there are right and left handed Christoffel symbols, and superficially they differ only in their Christoffel portion. They are discussed in more detail shortly.

For completeness, now define the 
``$ \; \tilde{;} \; $'' derivative as the reversal of the convention just given for the 
``$ \; ; \; $'' derivative. That means that all the tensor - Christoffel symbol positions are reversed for each term for each tensor index in the covariant derivative expression. For example,
\begin{equation}
g_{\mu \nu \, {\bf \tilde{;}} \, \gamma} = g_{\mu \nu , \gamma} 
- \left \{ \left. {}^{\: \alpha}_{\mu \gamma } \right ] \right. g_{\alpha \nu} 
- \left \{ \left. {}^{\: \alpha}_{\nu \gamma } \right ] \right. g_{\mu \alpha} 
\; ( \; \neq 0 ) 
\label{def.rev.cov.metric}
\end{equation}
By definition, the Christoffel symbols are always defined using a ``normal'' covariant derivative of the metric tensor, 
``$ \; ; \; $''. Clearly, tensor - Christoffel symbol positions in these expressions are determined {\it both} by the type of Christoffel symbol 
(``$ \{ ] $'' or ``$ [ \} $''), {\it and} the use of 
``$ \; ; \; $'' or 
``$ \; \tilde{;} \; $'' in the derivative.

With these facts established, the left handed Christoffel symbols and more general affine connection 
$ {}_{L} \Gamma^{\alpha}_{\mu \nu} $ will now be arbitrarily dropped, and the right handed cases used in what follows. However, it should also be noted that any action principle in this extended structure will eventually involve an added quaternion conjugate 
(``$ QC $'') term to keep the action real.\footnote{If a quaternion is written as a 
$ 2 \times 2 $ matrix, adding the 
$ QC $ term (the transpose of the complex conjugate) gives the same result as taking the trace of the matrix, which is what is done in Yang-Mills theory actions\cite{guidry}.} That quaternion conjugate term will tend to involve left handed forms to balance the right handed forms that are now being chosen in the first part of the action. Because of that, the basic left hand should not be suppressed or shortchanged overall, although a somewhat different type of right or left handedness will also arise at the next level to be examined. There, the right and left handed forms will be found necessarily to enter asymmetrically, but they should still tend to be balanced overall by corresponding opposite handed forms in the 
$ QC $ terms in the action.

For general quaternionic 
$ M_{\mu} $ and $ N_{\nu} $,
\begin{equation}
( M_{\mu} N_{\nu} )_{; \tau} \neq M_{\mu ; \tau} N_{\nu} + 
M_{\mu} N_{\nu ; \tau}
\label{def.no.product.rule}
\end{equation}
Additionally, contraction on tensor indices inside an already evaluated covariant derivative will not necessarily equal the covariant derivative of the contracted quantity. Examples such as these limit the usefulness of these gauge varying, generalized covariant derivatives outside of gauges in which quantities commute easily in products. However, in other cases, these covariant derivatives will still be helpful, and will be used. On the other hand, any genuine physics in this structure will require gauge invariant constructions, including gauge invariant covariant derivatives. Those will be developed later, and they will involve real, gauge invariant Christoffel symbols that thus avoid these limitations just noted.

\subsection{Weyl-Like Connections, Gauge Properties, and Curvatures}

Since right handed forms are now chosen, such as equation (\ref{def.right.chrst}), specialize equation (\ref{gbar.lr}) to 
$ \lambda = 1 $, or
\begin{equation}
\bar g_{\mu \nu} = g_{\mu \nu} \rho 
\label{gbar.r}
\end{equation}
Corresponding to that,
\begin{equation}
\bar g^{\mu \nu} = \rho^{-1} g^{\mu \nu} 
\label{gbar.i.r}
\end{equation}
These together with equation (\ref{def.right.chrst}) then give that
{\samepage 
\begin{eqnarray}
\bar{ \{ ^{\: \alpha}_{\mu \nu} ] } & = & 
\rho^{-1} \{ ^{\: \alpha}_{\mu \nu} ] \rho + 
{\textstyle{\frac{1}{2}}}\, \delta^{\alpha}_{\mu} \rho^{-1} \rho_{, \nu} + 
{\textstyle{\frac{1}{2}}}\, \delta^{\alpha}_{\nu} \rho^{-1} \rho_{, \mu} 
\nonumber \\
 &   & \; \; \; \; \; \; - {\textstyle{\frac{1}{2}}}\, \rho^{-1} 
g^{\alpha \tau} \! g_{\mu \nu} \, \rho_{, \tau} 
\nonumber \\
 & = & 
\rho^{-1} \{ ^{\: \alpha}_{\mu \nu} ] \rho + 
{\textstyle{\frac{1}{2}}}\, \delta^{\alpha}_{\mu} \rho^{-1} \rho_{, \nu} + 
{\textstyle{\frac{1}{2}}}\, \delta^{\alpha}_{\nu} \rho^{-1} \rho_{, \mu} 
\nonumber \\
 &   & \; \; \; \; \; \; - {\textstyle{\frac{1}{2}}}\, 
g^{\alpha \tau} \! g_{\mu \nu} \, \rho^{-1} \rho_{, \tau} 
\label{def.barchrst}
\end{eqnarray}}%
where the second equation follows from equation (\ref{gauge.inv.gpair}).

Weyl's original theory\cite{weyl.stm,adler.london} would now suggest an affine connection 
$ {}_{R} \Gamma^{\alpha}_{\mu \nu} $ (hereafter denoted by 
$ \Gamma^{\alpha}_{\mu \nu} $)
\begin{equation}
\Gamma^{\alpha}_{\mu \nu} = 
\{ ^{\: \alpha}_{\mu \nu} ] + 
\delta^{\alpha}_{\mu} v_{\nu} +
\delta^{\alpha}_{\nu} v_{\mu} - 
g^{\alpha \tau} \! g_{\mu \nu} v_{\tau}
\label{def.gamma.weyl}
\end{equation}
where $ v_{\mu} $ is the Weyl four vector, and
{\samepage 
\begin{eqnarray}
\bar{v}_{\mu} & = & \rho^{-1} \left ( v_{\mu} - 
{\textstyle{\frac{1}{2}}}\, 
\rho_{, \mu} \rho^{-1} \right ) \rho 
\nonumber \\
 & = & \rho^{-1} v_{\mu} \, \rho - {\textstyle{\frac{1}{2}}}\, 
\rho^{-1} \rho_{, \mu} 
\label{def.vbar.xtnd}
\end{eqnarray}}%
These give
\begin{equation}
\bar{\Gamma}^{\alpha}_{\mu \nu} = 
\rho^{-1} \Gamma^{\alpha}_{\mu \nu} \, \rho 
\label{gauge.gamma.weyl}
\end{equation}
as the quaternionic analog of the gauge invariance of the Weyl connection in his original geometry.

Equation (\ref{def.vbar.xtnd}) will indeed be adopted, but in place of equations (\ref{gauge.gamma.weyl}) and (\ref{def.gamma.weyl}), adopt
\begin{equation}
\bar{\Gamma}^{\alpha}_{\mu \nu} = 
\rho^{-1} \Gamma^{\alpha}_{\mu \nu} \, \rho + 
k \delta^{\alpha}_{\mu} \, \rho^{-1} \rho_{, \nu} 
\label{gauge.gamma.mixed.k}
\end{equation}
where $ k $ is a real constant, and
\begin{equation}
\Gamma^{\alpha}_{\mu \nu} = 
\{ ^{\: \alpha}_{\mu \nu} ] + 
n \delta^{\alpha}_{\mu} v_{\nu} +
\delta^{\alpha}_{\nu} v_{\mu} - 
g^{\alpha \tau} \! g_{\mu \nu} v_{\tau}
\label{def.gamma.mixed}
\end{equation}
Here $ n $ is also a real constant, and 
\begin{equation}
k = \frac{1 - n}{2}
\label{def.k.n}
\end{equation}
For 
$ n \neq 1 $, some amount of torsion appears, and furthermore, 
$ n = 0 $ gives a vanishing affine derivitive for the metric, a ``metric compatible'' case. Note that the cases 
$ n \neq 1 $ produce a form of Einstein's ``lambda transformation'' of the connection\cite{einstein.meaning} in equation (\ref{gauge.gamma.mixed.k}), which as he notes, leaves the curvature tensor invariant when the curvature tensor and lambda transformation involve real (or complex) quantities.

Now both of the connections 
$ \Gamma^{\alpha}_{\mu \nu} $ and 
$ \{ ^{\: \alpha}_{\mu \nu} ] $ have associated curvature tensors, but those have both a ``right handed'' and a ``left handed'' form themselves corresponding to use of normal or reversed covariant derivative conventions such as those used in 
``$ \; ; \; $'' or 
``$ \; \tilde{;} \; $'', irrespective of the right-left nature of the underlying connection used in them. Specifically,
\begin{equation}
{}_{R} B^{\gamma}_{\mu \tau \sigma} = 
\Gamma^{\gamma}_{\mu \sigma , \tau} - 
\Gamma^{\gamma}_{\mu \tau , \sigma} + 
\Gamma^{\gamma}_{\eta \tau} \Gamma^{\eta}_{\mu \sigma} - 
\Gamma^{\gamma}_{\eta \sigma} \Gamma^{\eta}_{\mu \tau} 
\label{def.b.r}
\end{equation}
\begin{equation}
{}_{L} B^{\gamma}_{\mu \tau \sigma} = 
\Gamma^{\gamma}_{\mu \sigma , \tau} - 
\Gamma^{\gamma}_{\mu \tau , \sigma} + 
\Gamma^{\eta}_{\mu \sigma} \Gamma^{\gamma}_{\eta \tau} - 
\Gamma^{\eta}_{\mu \tau} \Gamma^{\gamma}_{\eta \sigma} 
\label{def.b.l}
\end{equation}
\begin{equation}
{}_{R} R^{\gamma}_{\mu \tau \sigma} = 
\{ ^{\: \gamma}_{\mu \sigma} ]_{, \tau} - 
\{ ^{\: \gamma}_{\mu \tau} ]_{, \sigma} + 
\{ ^{\: \gamma}_{\eta \tau} ] \{ ^{\: \eta}_{\mu \sigma} ] - 
\{ ^{\: \gamma}_{\eta \sigma} ] \{ ^{\: \eta}_{\mu \tau} ] 
\label{def.r.r}
\end{equation}
and
\begin{equation}
{}_{L} R^{\gamma}_{\mu \tau \sigma} = 
\{ ^{\: \gamma}_{\mu \sigma} ]_{, \tau} - 
\{ ^{\: \gamma}_{\mu \tau} ]_{, \sigma} + 
\{ ^{\: \eta}_{\mu \sigma} ] \{ ^{\: \gamma}_{\eta \tau} ] - 
\{ ^{\: \eta}_{\mu \tau} ] \{ ^{\: \gamma}_{\eta \sigma} ] 
\label{def.r.l}
\end{equation}

When equation (\ref{gauge.gamma.mixed.k}) is substituted into equations (\ref{def.b.r}) and (\ref{def.b.l}) to obtain the gauge properties of those curvature tensors, generally neither curvature form transforms in a particularly neat manner by itself. Surprisingly however, the combination
\begin{equation}
B^{\gamma}_{\mu \tau \sigma} = 
\frac{k + 1}{2k} \, {}_{R} B^{\gamma}_{\mu \tau \sigma} + 
\frac{k - 1}{2k} \, {}_{L} B^{\gamma}_{\mu \tau \sigma} 
\label{def.b.neat.k}
\end{equation}
{\it does} have neat gauge transformation properties.\footnote{Reverse roles of 
$ {}_{R} \! B^{\gamma}_{\mu \tau \sigma} $ and 
$ {}_{L} \! B^{\gamma}_{\mu \tau \sigma} $ if left handed forms with 
$ \rho = 1$ and 
$ \lambda \ne 1 $ are adopted initially rather than right.} Specifically
\begin{equation}
\bar{B}^{\gamma}_{\mu \tau \sigma} = 
\rho^{-1} B^{\gamma}_{\mu \tau \sigma} \, \rho 
\label{def.b.neat.gauge}
\end{equation}
which is basically the same form as the gauge transformation of a Yang-Mills field in 
$ SU(2) $ gauge theory\cite{guidry}.

Furthermore, for real (or complex) quantities instead of quaternionic quantities, products like 
$ \Gamma^{\gamma}_{\eta \tau} \Gamma^{\eta}_{\mu \sigma} $ commute internally, and 
$ B^{\gamma}_{\mu \tau \sigma} $ clearly reduces to the usual curvature tensor form, and becomes gauge invariant like Weyl's curvature tensor\cite{weyl.stm}. Thus, it becomes the appropriate generalization of the curvature tensor in the quaternions. To facilitate its use in what follows, the coefficients in the definition of this tensor in equation (\ref{def.b.neat.k}) are given their own symbols,
\begin{equation}
k_{+} = \frac{k + 1}{2k}
\label{def.kplus}
\end{equation}
and
\begin{equation}
k_{-} = \frac{k - 1}{2k}
\label{def.kminus}
\end{equation}
Note that 
$ k_{+} + k_{-} = 1 $, and 
$ k_{+} - k_{-} = 1 / k $.

Now note that if one attempts to use a full Weyl connection analog by setting 
$ n = 1 $ in equation (\ref{def.gamma.mixed}), that causes 
$ k $ to become zero in equation (\ref{gauge.gamma.mixed.k}), and no suitable generalized curvature 
$ B^{\gamma}_{\mu \tau \sigma} $ emerges at all. Rather, as 
$ k $ approaches $ 0 $ in equation (\ref{def.b.neat.k}), the coefficients of 
$ {}_{R} B^{\gamma}_{\mu \tau \sigma} $ and 
$ {}_{L} B^{\gamma}_{\mu \tau \sigma} $ approach equal but opposite infinite values. Essentially, equation (\ref{def.b.neat.k}) must then be replaced by
\begin{equation}
B^{\gamma}_{\mu \tau \sigma} = 
{}_{R} \! B^{\gamma}_{\mu \tau \sigma} - 
{}_{L} \! B^{\gamma}_{\mu \tau \sigma} 
\label{def.b.neat.k.0}
\end{equation}
or any simple multiple of the right side of this equation, but that will lead to the disappearance of the derivatives of 
$ \Gamma^{\alpha}_{\mu \nu} $ from the result. Without the derivative terms, this quantity cannot reduce to anything at all like the curvature tensor of a Weyl geometry when quantities commute, reducing to zero instead. In other words, this structure actually discriminates against the exact quaternionic analog of Weyl's original theory\cite{weyl.stm}, and favors the cases which have some torsion. This is one primary reason the original Weyl connection must be unbalanced in order to generalize to the quaternions. Clearly 
$ n \neq 1 $ is necessary for a nontrivial structure.

Finally, the four vector 
$ v_{\mu} $ of equation (\ref{def.vbar.xtnd}) has its own directly associated Yang-Mills field tensor\cite{guidry}. The gauge properties give that
\begin{equation}
y_{\mu \nu} = 
v_{\nu ,\mu} - v_{\mu ,\nu} + 
2(v_{\nu} v_{\mu} - v_{\mu} v_{\nu}) 
\label{def.ym}
\end{equation}
gauge transforms just as 
$ B^{\gamma}_{\mu \tau \sigma} $ does, or
\begin{equation}
\bar{y}_{\mu \nu} = 
\rho^{-1} y_{\mu \nu} \, \rho 
\label{def.ym.gauge}
\end{equation}

Comparison of the transformation rules of equation (\ref{def.vbar.xtnd}) and the corresponding equation in Guidry then relates his 
$ A_{\mu} $ to 
$ v_{\mu} $ via
\begin{equation}
v_{\mu} = - \imath \frac{g}{2} A_{\mu}
\label{guidry.pot}
\end{equation}
where 
$ g $ is the coupling constant, and the 
$ - \imath $ is absorbed into the 
$ \sigma_{k} $ of equation (\ref{def.qk}) that are embedded within the 
$ A_{\mu} $, producing the imaginary basis quaternions 
$ Q_{k} $ in their place. Additionally, his 
$ U = \rho^{ - 1 } $ to complete the matchup with his 
$ SU(2) $ Yang-Mills theory.\footnote{Guidry's potential is totally imaginary in the quaternions, and as long as the gauge transformation 
$ \rho $ is unitary, or a real, nonzero constant times a unitary transformation, Guidry's potential remains totally imaginary.} As a bonus, one sees that the absorption of 
$ (g / 2) $ into 
$ A_{\mu} $ is what renders it dimensionless (if it is not already dimensionless), and suitable for this structure.

\subsection{The Makeup of the Curvature Tensor}

Equation (\ref{def.gamma.mixed}) can be written
\begin{equation}
\Gamma^{\alpha}_{\mu \nu} = 
\{ ^{\: \alpha}_{\mu \nu} ] + 
U^{\alpha}_{\mu \nu} 
\label{def.gamma.u}
\end{equation}
where
\begin{equation}
U^{\alpha}_{\mu \nu} = 
n \delta^{\alpha}_{\mu} v_{\nu} + 
\delta^{\alpha}_{\nu} v_{\mu} - 
g^{\alpha \tau} \! g_{\mu \nu} v_{\tau}
\label{def.u}
\end{equation}
Substituting these into equations (\ref{def.b.r}) and (\ref{def.b.l}) then gives the surprisingly neat results
\begin{equation}
{}_{R} B^{\gamma}_{\mu \tau \sigma} = 
{}_{R} R^{\gamma}_{\mu \tau \sigma} + 
U^{\gamma}_{\mu \sigma ; \tau} - 
U^{\gamma}_{\mu \tau ; \sigma} + 
U^{\gamma}_{\eta \tau} U^{\eta}_{\mu \sigma} - 
U^{\gamma}_{\eta \sigma} U^{\eta}_{\mu \tau} 
\label{def.b.r.u.r}
\end{equation}
and
\begin{equation}
{}_{L} B^{\gamma}_{\mu \tau \sigma} = 
{}_{L} R^{\gamma}_{\mu \tau \sigma} + 
U^{\gamma}_{\mu \sigma \, {\bf \tilde{;}} \, \tau} - 
U^{\gamma}_{\mu \tau \, {\bf \tilde{;}} \, \sigma} + 
U^{\eta}_{\mu \sigma} U^{\gamma}_{\eta \tau} - 
U^{\eta}_{\mu \tau} U^{\gamma}_{\eta \sigma} 
\label{def.b.r.u.l}
\end{equation}
These equations are one example (perhaps the best) in which the 
``$ \; ; \; $'' and 
``$ \; \tilde{;} \; $'' covariant derivatives give results that are both compact, and express useful information.

However, in order to proceed further with the evaluation of 
$ B^{\gamma}_{\mu \tau \sigma} $ via equations (\ref{def.b.neat.k}), (\ref{def.b.r.u.r}), and (\ref{def.b.r.u.l}), the covariant derivatives should be written out as partial derivatives and Christoffel symbol terms, and equation (\ref{def.u}) should be substituted into the result. Furthermore, any resulting partial derivatives of 
$ g_{\mu \nu} $ or $ g^{\mu \nu} $ should be evaluated using equations (\ref{def.right.cov.metric}) and (\ref{def.right.cov.imetric}) to substitute terms with Christoffel symbols in place of the partial derivative terms. In practice, the combination 
$ ( g^{\gamma \eta} g_{\mu \sigma} )_{, \tau} $ always appears as a unit, and can be eliminated using
{\samepage 
\begin{eqnarray}
( g^{\gamma \eta} g_{\mu \sigma} )_{, \tau} & = & 
g^{\gamma \eta} g_{\alpha \sigma} \{ ^{\: \alpha}_{\mu \tau} ] + 
g^{\gamma \eta} g_{\mu \alpha} \{ ^{\: \alpha}_{\sigma \tau} ] 
\nonumber \\
 & & - \{ ^{\: \gamma}_{\alpha \tau} ] g^{\alpha \eta} g_{\mu \sigma} - 
\{ ^{\: \eta}_{\alpha \tau} ] g^{\gamma \alpha} g_{\mu \sigma} 
\label{def.cov.gunit}
\end{eqnarray}}%
keeping equation (\ref{gauge.inv.gpair}) in mind for the result. The fact that the 
$ g^{\gamma \eta} g_{\mu \sigma} $ terms are real then allows equation (\ref{def.cov.gunit}) to have more than one valid form simply by varying the position of such terms in its products. However, the same form should consistently be chosen internally throughout evaluation of either one of the {\it separate} tensors in the pair 
$ {}_{R} B^{\gamma}_{\mu \tau \sigma} $ or 
$ {}_{L} B^{\gamma}_{\mu \tau \sigma} $ to avoid possibly encountering extraneous terms that should evaluate to zero with some effort, but are more easily avoided from the outset.\footnote{Extraneous terms can be a problem particularly when verifying overall gauge transformation properties and complete gauge balancing of the sum of all terms after the substitutions.} Additionally, the full expression of equation (\ref{def.cov.gunit}) itself should be real, and could be moved around as a unit in products in its containing equation if necessary. However, all this flexibility leads to more than one expansion of 
$ B^{\gamma}_{\mu \tau \sigma} $ in gauge varying quantities like 
$ v_{\mu} $, although all the expansions are equivalent, and all will lead to the same, unique, gauge invariant result in what follows. Since the gauge invariant result contains any real physics, its uniqueness is what is important.

The result of the above substitutions and expansions gives
{\samepage 
\begin{eqnarray}
B^{\gamma}_{\mu \tau \sigma} & = & 
k_{+} \, {}_{R} \! R^{\gamma}_{\mu \tau \sigma} + 
k_{-} \, {}_{L} \! R^{\gamma}_{\mu \tau \sigma} 
\nonumber \\
 & & + \left ( k_{+} \, v_{\mu ; \tau} + 
k_{-} \, v_{\mu \, {\bf \tilde{;}} \, \tau} 
\right ) \delta^{\gamma}_{\sigma} 
\nonumber \\
 & & - \left ( k_{+} \, v_{\mu ; \sigma} + 
k_{-} \, v_{\mu \, {\bf \tilde{;}} \, \sigma} 
\right ) \delta^{\gamma}_{\tau} 
\nonumber \\
 & & - \left ( k_{+} \, v_{\eta ; \tau} + 
k_{-} \, v_{\eta \, {\bf \tilde{;}} \, \tau} 
\right ) g^{\eta \gamma} g_{\mu \sigma} 
\nonumber \\
 & & + \left ( k_{+} \, v_{\eta ; \sigma} + 
k_{-} \, v_{\eta \, {\bf \tilde{;}} \, \sigma} 
\right ) g^{\eta \gamma} g_{\mu \tau} 
\nonumber \\
 & & + \frac{1}{k} \left ( \left [ v_{\eta} , 
\{ ^{\: \gamma}_{\alpha \tau} ] \, \right ] 
g^{\eta \alpha} g_{\mu \sigma} - \left [ v_{\eta} , 
\{ ^{\: \gamma}_{\alpha \sigma} ] \, \right ] 
g^{\eta \alpha} g_{\mu \tau} \right ) 
\nonumber \\
 & & + \left ( k_{+} \, v_{\sigma} v_{\mu} + 
k_{-} \, v_{\mu} v_{\sigma} \right ) 
\delta^{\gamma}_{\tau} 
\nonumber \\
 & & - \left ( k_{+} \, v_{\tau} v_{\mu} + 
k_{-} \, v_{\mu} v_{\tau} \right ) 
\delta^{\gamma}_{\sigma} 
\nonumber \\
 & & - v_{\eta} v_{\beta} g^{\eta \beta} \left ( 
g_{\mu \sigma} \delta^{\gamma}_{\tau} - 
g_{\mu \tau} \delta^{\gamma}_{\sigma} \right ) 
\nonumber \\
 & & + \left ( k_{+} \, v_{\alpha} v_{\tau} + 
k_{-} \, v_{\tau} v_{\alpha} \right ) 
g^{\alpha \gamma} g_{\mu \sigma} 
\nonumber \\
 & & - \left ( k_{+} \, v_{\alpha} v_{\sigma} + 
k_{-} \, v_{\sigma} v_{\alpha} \right ) 
g^{\alpha \gamma} g_{\mu \tau} 
\nonumber \\
 & & + n \left [ k_{+} \left ( v_{\sigma ; \tau} - 
v_{\tau ; \sigma} \right ) + k_{-} \left ( 
v_{\sigma \, {\bf \tilde{;}} \, \tau} - 
v_{\tau \, {\bf \tilde{;}} \, \sigma} \right ) 
\right ] \delta^{\gamma}_{\mu} 
\nonumber \\
& & - \frac{n}{k} \left ( \left [ v_{\sigma} , 
\{ ^{\: \gamma}_{\mu \tau} ] \, \right ] - 
\left [ v_{\tau} , \{ ^{\: \gamma}_{\mu \sigma} ] 
\, \right ] \right ) 
\nonumber \\
& & - \frac{n^{2}}{k} \left ( v_{\sigma} 
v_{\tau} - v_{\tau} v_{\sigma} \right ) 
\delta^{\gamma}_{\mu} 
\nonumber \\
& & - \frac{n}{k} \left [ \left ( v_{\mu} v_{\tau} - 
v_{\tau} v_{\mu} \right ) \delta^{\gamma}_{\sigma} - 
\left ( v_{\mu} v_{\sigma} - v_{\sigma} v_{\mu} 
\right ) \delta^{\gamma}_{\tau} \right ] 
\nonumber \\
& & + \frac{n}{k} \left ( v_{\alpha} v_{\tau} - 
v_{\tau} v_{\alpha} \right ) 
g^{\alpha \gamma} g_{\mu \sigma} 
\nonumber \\
 & & - \frac{n}{k} \left ( v_{\alpha} v_{\sigma} - 
v_{\sigma} v_{\alpha} \right ) 
g^{\alpha \gamma} g_{\mu \tau} 
\label{def.expanded.b}
\end{eqnarray}}%
where the 
``$ [ \; , \; ] $'' terms are conventional commutators. Those commutators will clearly vanish in gauges in which 
$ \{ ^{\: \gamma}_{\alpha \sigma} ] $ is real.

Now 
$ B^{\gamma}_{\mu \tau \sigma} $ can be contracted to give
\begin{equation}
B_{\mu \tau} = B^{\omega}_{\mu \tau \omega} 
\label{def.contract.b}
\end{equation}
and clearly equation (\ref{def.b.neat.gauge}) gives that
\begin{equation}
\bar{B}_{\mu \tau} = 
\rho^{-1} B_{\mu \tau} \, \rho 
\label{def.contract.b.gauge}
\end{equation}
The similarity between this equation and equation (\ref{def.ym.gauge}) might then raise expectations that the antisymmetric part of 
$ B_{\mu \tau} $ will be proportional to 
$ y_{\mu \tau} $ once some method is found for contracting the right side of equation (\ref{def.expanded.b}) to a neat expression. However, this will generally not be quite true. A check reveals that the antisymmetric part of 
$ k_{+} \, {}_{R} \! R^{\omega}_{\mu \tau \omega} + 
k_{-} \, {}_{L} \! R^{\omega}_{\mu \tau \omega} $
equals 
$ \{ 1 - [ 1 / ( 4 k ) ] \} [ \gamma^{-1} \gamma_{, \mu} , 
\gamma^{-1} \gamma_{, \tau} ] $ 
where $ \gamma $ is the gauge function in 
$ g_{\mu \nu} = \tilde{g}_{\mu \nu} \gamma $. Since this commutator does not generally vanish, then the antisymmetric part of 
$ k_{+} \, {}_{R} \! R^{\omega}_{\mu \tau \omega} + 
k_{-} \, {}_{L} \! R^{\omega}_{\mu \tau \omega} $ 
is generally not zero, and that antisymmetric component must be gauge balanced elsewhere by antisymmetric terms, even though it vanishes in gauges in which 
$ g_{\mu \nu} $ is real, and also when 
$ \gamma $ remains in the complex plane. However, the obvious exception is the case 
$ k = 1 / 4 $, which causes the term to vanish even when the commutator is nonzero. That special value of 
$ k $ corresponds to 
$ n = 1 / 2 $. The consequences of all this will become clearer in the next section where a simple method is developed to express contractions of the expansion of 
$ B^{\gamma}_{\mu \tau \sigma} $ given by equation (\ref{def.expanded.b}).

Finally define the scalar curvature
\begin{equation}
B = B_{\mu \tau} g^{\mu \tau} 
\label{def.scalar.b}
\end{equation}
Since
\begin{equation}
\bar{g}^{\mu \tau} = 
\rho^{-1} g^{\mu \tau} 
\label{def.asym.imetric.gauge}
\end{equation}
then equations (\ref{def.contract.b.gauge}) and (\ref{def.scalar.b}) give
\begin{equation}
\bar{B} = 
\rho^{-1} B 
\label{def.scalar.b.gauge}
\end{equation}
Thus $ B $ is the key quantity needed to define gauge invariant variables. As conceived by Weyl and Eddington\cite{weyl.stm,eddington.mtr}, it is basically an intrinsic yardstick provided by the spacetime structure itself to reduce equations to dimensionless, gauge invariant quantities that can correspond to actual physics. It allows the structure to measure itself. For this purpose, it is assumed to be nonzero.

\subsection{Gauge Invariant Variables and Their Fundamental Identity}

Define
\begin{equation}
\hat g_{\mu \nu} = g_{\mu \nu} (B/C)
\label{def.qghat}
\end{equation}
and its inverse
\begin{equation}
\hat g^{\mu \nu} = C B^{-1} g^{\mu \nu} 
\label{def.qighat}
\end{equation}
where 
$ C $ is a constant to be explained shortly. Using equations (\ref{gbar.r}) and (\ref{def.scalar.b.gauge}), these are seen to have the same value in all gauges. They are gauge invariant forms of the metric and its inverse. As such, they are immediately assumed to be real, giving a real metric tensor that can be used in physics. The constant 
$ C $ is necessary if all quantities are real, and the scalar curvature 
$ B < 0 \; $. Then in order to keep the signatures of 
$ g_{\mu \nu} $ and 
$ \hat g_{\mu \nu} $ from being opposite, 
$ C = - 1 $ would be appropriate.\footnote{My thanks to Daniel Galehouse for correctly insisting that cases with negative values of 
$ B $, thus implying negative 
$ C $, are legitimate\cite{galehouse.ijtp.1,galehouse.ijtp.2}.} Thus having established its right to exist, 
$ C $ is retained as a (real, dimensionless) constant in 
general.

There is a gauge invariant 
$ \hat{ \{ ^{\: \alpha}_{\mu \nu} \} } $ based on the real 
$ \hat g_{\mu \nu} $, and it is real,
\begin{equation}
\hat{ \{ ^{\: \alpha}_{\mu \nu} \} } = 
\frac{1}{2} \; \hat g^{\alpha \tau} ( \hat g_{\mu \tau , \nu} + 
\hat g_{\nu \tau , \mu} - \hat g_{\mu \nu , \tau} ) 
\label{def.hat.chrst}
\end{equation}
without a right-left nature any longer. The covariant derivative with respect to it is indicated by 
``$ {}_{\|} $'', and it is now quite well behaved, including obeying the product rule since 
$ \hat{ \{ ^{\: \alpha}_{\mu \nu} \} } $ commutes with everything. Both the 
``$ \; \tilde{;} \; $'' and the 
``$ \; ; \; $'' covariant derivative conventions will reduce to it. If equations (\ref{def.qghat}) and (\ref{def.qighat}) are substituted into equation (\ref{def.hat.chrst}), the result gives
{\samepage 
\begin{eqnarray}
\hat{ \{ ^{\: \alpha}_{\mu \nu} \} } & = & 
B^{-1} \{ ^{\: \alpha}_{\mu \nu} ] B + 
{\textstyle{\frac{1}{2}}}\, \delta^{\alpha}_{\mu} B^{-1} B_{, \nu} + 
{\textstyle{\frac{1}{2}}}\, \delta^{\alpha}_{\nu} B^{-1} B_{, \mu} 
\nonumber \\
 &   & \; \; \; \; \; \; - {\textstyle{\frac{1}{2}}}\, B^{-1} 
g^{\alpha \tau} \! g_{\mu \nu} \, B_{, \tau} 
\nonumber \\
 & = & 
B^{-1} \{ ^{\: \alpha}_{\mu \nu} ] B + 
{\textstyle{\frac{1}{2}}}\, \delta^{\alpha}_{\mu} B^{-1} B_{, \nu} + 
{\textstyle{\frac{1}{2}}}\, \delta^{\alpha}_{\nu} B^{-1} B_{, \mu} 
\nonumber \\
 &   & \; \; \; \; \; \; - {\textstyle{\frac{1}{2}}}\, 
g^{\alpha \tau} \! g_{\mu \nu} \, B^{-1} B_{, \tau} 
\label{def.expand.hat.chrst}
\end{eqnarray}}%

These real, commuting Christoffel symbols now give us a normal, real, gauge invariant Riemannian geometry on which we can impose a form of General Relativity. They define a Riemann curvature tensor 
$ \hat R^{\gamma}_{\mu \tau \sigma} $, and both the conventions of 
$ {}_{R} R^{\gamma}_{\mu \tau \sigma} $ in equation (\ref{def.r.r}), and of 
$ {}_{L} R^{\gamma}_{\mu \tau \sigma} $ in equation (\ref{def.r.l}) reduce to it. There is a (now symmetric) 
$ \hat R_{\mu \tau} = \hat R^{\omega}_{\mu \tau \omega} $ , and a scalar 
$ \hat R = \hat g^{\mu \tau} \hat R_{\mu \tau} $.

The gauge invariant Weyl vector is
{\samepage 
\begin{eqnarray}
\hat v_{\mu} & = & B^{-1} \left ( v_{\mu} - 
{\textstyle{\frac{1}{2}}}\, 
B_{, \mu} B^{-1} \right ) B 
\nonumber \\
 & = & B^{-1} v_{\mu} B - {\textstyle{\frac{1}{2}}}\, 
B^{-1} B_{, \mu} 
\label{def.qvhat}
\end{eqnarray}}%
which is fully quaternionic generally. Then in analogy to equation (\ref{def.gamma.mixed}), define the gauge invariant
\begin{equation}
\hat \Gamma^{\alpha}_{\mu \nu} = 
\hat{\{ ^{\: \alpha}_{\mu \nu} \} } + 
n \delta^{\alpha}_{\mu} \hat v_{\nu} + 
\delta^{\alpha}_{\nu} \hat v_{\mu} - 
\hat g^{\alpha \tau} \! \hat g_{\mu \nu} \hat v_{\tau}
\label{def.gamma.hat.mixed}
\end{equation}
Note that since 
$ \hat \Gamma^{\alpha}_{\mu \nu} $ is fully quaternionic, the full affine derivative of a quantity using 
$ \hat \Gamma^{\alpha}_{\mu \nu} $ is {\it not} as well behaved as the covariant derivative using only the real 
$ \hat{ \{ ^{\: \alpha}_{\mu \nu} \} } $.

Now substituting into equation (\ref{def.gamma.hat.mixed}) from equations (\ref{def.expand.hat.chrst}) and (\ref{def.qvhat}), and using equation (\ref{def.gamma.mixed}), one sees
\begin{equation}
\hat \Gamma^{\alpha}_{\mu \nu} = 
B^{-1} \Gamma^{\alpha}_{\mu \nu} \, B + 
k \delta^{\alpha}_{\mu} \, B^{-1} B_{, \nu} 
\label{gamma.hat.gamma}
\end{equation}
But this is exactly the same form as a gauge transformation on 
$ \Gamma^{\alpha}_{\mu \nu} $ as defined in equation (\ref{gauge.gamma.mixed.k}). Thus, if one defines 
$ {}_{R} \hat B^{\gamma}_{\mu \tau \sigma} $ and 
$ {}_{L} \hat B^{\gamma}_{\mu \tau \sigma} $ using 
$ \hat \Gamma^{\alpha}_{\mu \nu} $ in full analogy to the use of 
$ \Gamma^{\alpha}_{\mu \nu} $ in 
$ {}_{R} B^{\gamma}_{\mu \tau \sigma} $ and 
$ {}_{L} B^{\gamma}_{\mu \tau \sigma} $ of equations (\ref{def.b.r}) and (\ref{def.b.l}), the result gives finally that
{\samepage 
\begin{eqnarray}
\hat B^{\gamma}_{\mu \tau \sigma} & = & 
k_{+} \, {}_{R} \! \hat B^{\gamma}_{\mu \tau \sigma} + 
k_{-} \, {}_{L} \! \hat B^{\gamma}_{\mu \tau \sigma} 
\nonumber \\
 & = & B^{-1} B^{\gamma}_{\mu \tau \sigma} B 
\label{def.bhat}
\end{eqnarray}}%

This can be expanded just like equation (\ref{def.expanded.b}), but now with so many quantities real, the much simpler result is
{\samepage 
\begin{eqnarray}
\hat B^{\gamma}_{\mu \tau \sigma} & = & 
\hat R^{\gamma}_{\mu \tau \sigma} + \hat v_{\mu \| \tau} 
\delta^{\gamma}_{\sigma} - 
\hat v_{\mu \| \sigma} 
\delta^{\gamma}_{\tau} 
\nonumber \\
 & & - \hat v_{\eta \| \tau} 
\hat g^{\eta \gamma} \hat g_{\mu \sigma} + 
\hat v_{\eta \| \sigma} 
\hat g^{\eta \gamma} \hat g_{\mu \tau} 
\nonumber \\
 & & + \left ( k_{+} \, \hat v_{\sigma} \hat v_{\mu} + 
k_{-} \, \hat v_{\mu} \hat v_{\sigma} \right ) 
\delta^{\gamma}_{\tau} 
\nonumber \\
 & & - \left ( k_{+} \, \hat v_{\tau} \hat v_{\mu} + 
k_{-} \, \hat v_{\mu} \hat v_{\tau} \right ) 
\delta^{\gamma}_{\sigma} 
\nonumber \\
 & & - \hat v_{\eta} \hat v_{\beta} \hat g^{\eta \beta} 
\left ( \hat g_{\mu \sigma} 
\delta^{\gamma}_{\tau} - \hat g_{\mu \tau} 
\delta^{\gamma}_{\sigma} \right ) 
\nonumber \\
 & & + \left ( k_{+} \, \hat v_{\alpha} \hat v_{\tau} + 
k_{-} \, \hat v_{\tau} \hat v_{\alpha} \right ) 
\hat g^{\alpha \gamma} \hat g_{\mu \sigma} 
\nonumber \\
 & & - \left ( k_{+} \, \hat v_{\alpha} \hat v_{\sigma} 
+ k_{-} \, \hat v_{\sigma} \hat v_{\alpha} \right ) 
\hat g^{\alpha \gamma} \hat g_{\mu \tau} 
\nonumber \\
 & & + n \left ( \hat v_{\sigma \| \tau} - 
\hat v_{\tau \| \sigma} \right ) \delta^{\gamma}_{\mu} 
\nonumber \\
 & & - \frac{n^{2}}{k} \left ( \hat v_{\sigma} 
\hat v_{\tau} - \hat v_{\tau} \hat v_{\sigma} \right ) 
\delta^{\gamma}_{\mu} 
\nonumber \\
 & & - \frac{n}{k} \left [ \left ( \hat v_{\mu} 
\hat v_{\tau} - \hat v_{\tau} \hat v_{\mu} \right ) 
\delta^{\gamma}_{\sigma} - \left ( \hat v_{\mu} 
\hat v_{\sigma} - \hat v_{\sigma} \hat v_{\mu} \right ) 
\delta^{\gamma}_{\tau} \right ] 
\nonumber \\
 & & + \frac{n}{k} \left ( \hat v_{\alpha} 
\hat v_{\tau} - \hat v_{\tau} \hat v_{\alpha} \right ) 
\hat g^{\alpha \gamma} \hat g_{\mu \sigma} 
\nonumber \\
 & & - \frac{n}{k} \left ( \hat v_{\alpha} 
\hat v_{\sigma} - \hat v_{\sigma} \hat v_{\alpha} \right ) 
\hat g^{\alpha \gamma} \hat g_{\mu \tau} 
\label{def.expanded.bhat}
\end{eqnarray}}%
Much of the right-left distinction of equation (\ref{def.expanded.b}), along with the commutators, is now gone. The main left-right distinction remaining is in the terms involving products of 
$ \hat v_{\mu} $, because that quantity is fully quaternionic still.

Now using equations (\ref{def.k.n}), (\ref{def.kplus}), and (\ref{def.kminus}), equation (\ref{def.expanded.bhat}) and equation (\ref{def.bhat}) then contract to give
{\samepage 
\begin{eqnarray}
\hat B_{\mu \tau} & = & \hat B^{\omega}_{\mu \tau \omega} 
\nonumber \\
 & = & B^{-1} B_{\mu \tau} B 
\nonumber \\
 & = & \hat R_{\mu \tau} + 
\left ( \hat v_{\mu \| \tau} + \hat v_{\tau \| \mu} \right ) + 
\hat v^{\alpha}_{\; \; \| \alpha} \hat g_{\mu \tau} 
\nonumber \\
 & & - \left ( \hat v_{\mu} \hat v_{\tau} + 
\hat v_{\tau} \hat v_{\mu} \right ) + 
2 \hat v^{\alpha} \hat v_{\alpha} 
\hat g_{\mu \tau} 
\nonumber \\
 & & + \left ( 1 + n \right ) \left ( \hat v_{\mu \| \tau} - 
\hat v_{\tau \| \mu} \right ) 
\nonumber \\
 & & + \frac{4 - 4 n - 2 n^{2}}{1 - n} 
\left ( \hat v_{\mu} \hat v_{\tau} - 
\hat v_{\tau} \hat v_{\mu} \right ) 
\label{def.expanded.contract.bhat}
\end{eqnarray}}%
where the symmetric and antisymmetric parts have been clearly separated with the antisymmetric part all on the last two lines. Because 
$ \hat v_{\mu \| \tau} - \hat v_{\tau \| \mu} = \hat v_{\mu , \tau} - 
\hat v_{\tau , \mu} $, 
that antisymmetric part is
{\samepage 
\begin{eqnarray}
- \hat w_{\mu \tau} & = & 
\left ( 1 + n \right ) \left ( \hat v_{\mu \| \tau} - 
\hat v_{\tau \| \mu} \right ) 
\nonumber \\
 &   & \; \; \; \; \; \; + \frac{4 - 4 n - 2 n^{2}}{1 - n} 
\left ( \hat v_{\mu} \hat v_{\tau} - 
\hat v_{\tau} \hat v_{\mu} \right ) 
\nonumber \\
 & = & 
\left ( 1 + n \right ) \left ( \hat v_{\mu , \tau} - 
\hat v_{\tau , \mu} \right ) 
\nonumber \\
 &   & \; \; \; \; \; \; + \frac{4 - 4 n - 2 n^{2}}{1 - n} 
\left ( \hat v_{\mu} \hat v_{\tau} - 
\hat v_{\tau} \hat v_{\mu} \right ) 
\nonumber \\
 & = & 
- \left ( 1 + n \right ) \left [ \hat v_{\tau , \mu} - 
\hat v_{\mu , \tau} + 2 \left ( \hat v_{\tau} \hat v_{\mu} - 
\hat v_{\mu} \hat v_{\tau} \right ) \right. 
\nonumber \\
 &   & \; \; \; \; \; \; + \left. \frac{2 - 4 n}{1 - n^2} 
\left ( \hat v_{\tau} \hat v_{\mu} - 
\hat v_{\mu} \hat v_{\tau} \right ) \right ] 
\nonumber \\
 & = & 
- \left ( 1 + n \right ) \left [ \hat y_{\mu \tau} 
+ \frac{2 - 4 n}{1 - n^2} 
\left ( \hat v_{\tau} \hat v_{\mu} - 
\hat v_{\mu} \hat v_{\tau} \right ) \right ] 
\label{def.w.hat}
\end{eqnarray}}%
where
{\samepage 
\begin{eqnarray}
\hat y_{\mu \tau} & = & 
\hat v_{\tau , \mu} - \hat v_{\mu , \tau} + 
2 (\hat v_{\tau} \hat v_{\mu} - 
\hat v_{\mu} \hat v_{\tau}) 
\nonumber \\
 & = & B^{-1} y_{\mu \tau} B 
\label{def.yhat}
\end{eqnarray}}%
by equations (\ref{def.ym}) and (\ref{def.ym.gauge}), since equation (\ref{def.qvhat}) has the same form as the gauge transformation of 
$ v_{\mu} $ in equation (\ref{def.vbar.xtnd}). Additionally, the distinct alternate contraction of 
$ \hat B^{\gamma}_{\mu \tau \sigma} $ gives
\begin{equation}
\hat B^{\gamma}_{\gamma \tau \sigma} = 
4 \left [ n \hat y_{\tau \sigma} + 
\frac{1 - 2 n}{1 - n} 
\left ( \hat v_{\sigma} \hat v_{\tau} - 
\hat v_{\tau} \hat v_{\sigma} \right ) \right ] 
\label{def.expanded.alt.contract.bhat}
\end{equation}
This contraction of the curvature tensor is also important in Weyl's original theory\cite{weyl.stm,eddington.mtr}.

We have
\begin{equation}
\hat w_{\mu \nu} = 
\left ( 1 + n \right ) \left [ \hat y_{\mu \nu} + 
\frac{2 - 4 n}{1 - n^2} 
\left ( \hat v_{\nu} \hat v_{\mu} - 
\hat v_{\mu} \hat v_{\nu} \right ) \right ] 
\label{def.w.hat.tail}
\end{equation}
For $ n \neq 1 / 2 $, the quantity 
$ \hat w_{\mu \nu} $ appears to have a tail on it in addition to 
$ \hat y_{\mu \nu} $. This was anticipated above when the antisymmetric part of 
$ \, k_{+} \, {}_{R} \! R^{\omega}_{\mu \tau \omega} + 
k_{-} \, {}_{L} \! R^{\omega}_{\mu \tau \omega} $ was noted to require additional antisymmetric terms to gauge balance it unless 
$ n = 1 / 2 $. This is the form those extra terms take in 
$ \hat w_{\mu \nu} $.

Given that 
$ B = B_{\mu \tau} g^{\mu \tau} $, 
$ \; \hat g^{\mu \tau} = C B^{-1} g^{\mu \tau} $, and 
$ \hat B_{\mu \tau} = B^{-1} B_{\mu \tau} B $, then
{\samepage 
\begin{eqnarray}
\hat B & = & \hat B_{\mu \tau} \hat g^{\mu \tau} 
\nonumber \\
 & = & C B^{-1} B_{\mu \tau} g^{\mu \tau} 
\nonumber \\
 & = & C 
\label{def.scalar.bhat}
\end{eqnarray}}%
This is a fundamental, kinematic identity the gauge invariant variables must satisfy by virtue of their definitions, and the geometry's kinematics. Substituting from equations (\ref{def.expanded.contract.bhat}), and (\ref{def.w.hat}) for the expansion of 
$ \hat B_{\mu \tau} $, equation (\ref{def.scalar.bhat}) becomes
\begin{equation}
\hat R + 6 \hat v^{\mu}_{\; \; \| \mu} + 6 \hat v^{\mu} 
\hat v_{\mu} 
 = C
\label{def.identity.q}
\end{equation}
As an additional point that will be of use later, note that if all the derivatives 
($ {}_{, \mu} $) and coordinates in this result are reexpressed using (locally Lorentzian) standard lab coordinates 
$ x^{\mu}_{LAB} = x^{\mu} / \sqrt{ b_{0} } $ which have standard dimensions or units, then equation (\ref{def.identity.q}) becomes
\begin{equation}
\hat R + 6 \hat v^{\mu}_{\; \; \| \mu} + 6 \hat v^{\mu} 
\hat v_{\mu} 
 = b_{0} C
\label{def.identity.q.b0}
\end{equation}
The second term in equation (\ref{def.qvhat}) is the most useful one in understanding how 
$ \hat v_{\mu} $ is affected by this reversion to lab coordinates with dimensions.

Obviously the case 
$ n = 1 / 2 $ gives the simplest expression for 
$ \hat w_{\mu \nu} $ in equation (\ref{def.w.hat.tail}). Furthermore, for 
$ n = 1 / 2 $, and only for this value of 
$ n $, 
$ \hat B^{\gamma}_{\gamma \tau \sigma} $ given by equation (\ref{def.expanded.alt.contract.bhat}) is proportional to 
$ \hat w_{\tau \sigma} $ given by equation (\ref{def.w.hat}). That proportionality is a property that is true in Weyl's original theory\cite{weyl.stm}, so the case 
$ n = 1 / 2 $ is the only case that matches that property of Weyl's original theory (which had 
$ n = 1 $, a value not allowed in this model). One can fairly say that the 
$ n = 1 / 2 $ case is the closest one can get to Weyl's original model when generalizing to the quaternions, and it emphasizes the standard 
$ SU(2) $ Yang-Mills field (with optional additional real part) as a unique antisymmetric tensor in the structure, thereby unifying it with the framework of Einstein's Riemannian General Relativity naturally.

The cost of all these simplifications introduced by choosing 
$ n = 1 / 2 $, is that the structure now has an equal mix of Weyl's nonmetricity\cite{weyl.stm,eddington.mtr} with torsion, rather than insisting on just nonmetricity or torsion alone. That may seem unusual for a model in which the non-Riemannian behavior is primarily based on a Weyl-like four vector. Nevertheless, it does achieve a notable reduction in the complexity of the results. It's interesting to see that quaternionic curvatures in this model seem not only to reject the quaternionic generalization of the pure Weyl model, as noted after equation (\ref{def.b.neat.k.0}), but that they also preferentially select this case with an equal balance of torsion and nonmetricity. That preference is expressed by the overall simplicity of this case. No such preferential selection between torsion and nonmetricity would appear with purely real or complex gauges and curvatures.

However, note that it is {\it also} true that effective nonmetricity may not vanish even when 
$ n = 0 $, even though the full affine derivative of 
$ \hat g_{\mu \nu} $ using connection 
$ \hat \Gamma^{\alpha}_{\mu \nu} $ vanishes then, implying metric compatibility. This effective nonmetricity can be seen by looking at equation (\ref{def.expanded.alt.contract.bhat}) in the 
$ n = 0 $ case, and noting that the change in length of a vector transported using the full affine connection around a closed loop involves this quantity\cite{eddington.mtr}. This may still be nonzero even in the 
$ n = 0 $ case here, because 
$ \hat v_{\mu} $ is fully quaternionic. Thus, it appears that there may be no quaternionic models in this family which are completely devoid of all aspects of nonmetricity. This result appears to follow from the fact that covariant and contravariant vectors interact with the affine connection on opposite sides of the (quaternionic) connection, and the length of a vector is a contraction of a covariant and a contravariant vector. To put this another way, the affine derivative using the full 
$ \hat \Gamma^{\alpha}_{\mu \nu} $, no longer obeys the product rule of differentiation because 
$ \hat \Gamma^{\alpha}_{\mu \nu} $ is quaternionic, not real. Thus, the calculations of Weyl and Eddington\cite{weyl.stm,eddington.mtr} which would give the change in a parallel transported vector's length around a closed loop in terms of the affine derivative of the metric (which vanishes given metric compatibility), would no longer be completely valid.

\section{The Original, Real Variable Weyl Action}
\label{discuss}

At this point, the kinematical framework of a quaternionic Weyl-like geometry is in place. The use of gauge invariant variables allows definition of a real metric tensor suitable for General Relativity, and also produces an 
$ SU(2) $ Yang-Mills field, with an added possible real component as well, since the quaternions also have a real component. Quaternions are visualized as a four dimensional Euclidian space which is 
$ SU(2) \times SU(2) $, or 
$ SO(4) $.

Beyond this, the framework is so far a general, blank slate, since an action principle should define a particular dynamics to proceed further. In that regard, it is perfectly legitimate to phrase the action in terms of the gauge invariant variables, and indeed, it may be the only easy way to formulate an action without having to keep track of left and right handed parts, and other complications. However, if the gauge invariant variables are used, equation (\ref{def.identity.q}) must be included in the action as a constraint with a Lagrange multiplier, since the gauge invariant variables are not independent. That leads to some surprises.

To best illustrate this, if everything is restricted to the real numbers, and the original Weyl action is translated into gauge invariant variables\cite{rankin.ijtp}, then generalized slightly, it becomes for this structure
{\samepage 
\begin{eqnarray}
I & = & \int { \sqrt{-\hat g}\, \left [ 
\left ( \hat R - 2 \sigma \right ) - 
\frac{1}{2} j^{2} \left ( \hat y^{\mu \nu} 
\hat y_{\mu \nu} \right ) + 6 \hat v^{\mu} 
\hat v_{\mu} \right. } 
\nonumber \\
 &   & + \left. \left ( \hat R + 6 
\hat v^{\mu}_{\; \; \| \mu} + 6 \hat v^{\mu} 
\hat v_{\mu} - C \right ) \hat \beta \right ] d^4x 
\label{orig.gi.weyl.action}
\end{eqnarray}}%
where 
$ j^{2} $ is a dimensionless gravitational constant, and 
$ \hat \beta $ is the Lagrange multiplier. The fact that this is a modified, unbalanced Weyl structure does not affect this form. The contraction of the gauge invariant torsion with itself has the value 
$ ( 3 / 8) \hat v^{\mu} \hat v_{\mu} $, but additional terms based separately on that are not included here. For 
$ \sigma = \frac{1}{4} $ and 
$ C = 1 $, this is the original Weyl action\cite{ rankin.ijtp,rankin.caqg,rankin.9404023}, a fact which also can be seen by substituting for 
$ \hat R $ in it everywhere by using the constraint, discarding any total divergence terms, and expanding surviving terms into the unhatted Weyl variables. The Yang-Mills tensor reduces to the curl of the Weyl vector here, or essentially what Weyl called the electromagnetic field. In practice, the constraint of equation (\ref{def.identity.q}) is used to determine 
$ B $, given 
$ \hat g_{\mu \nu} $, and a value of 
$ v_{\mu} $ in any particular gauge\cite{rankin.ijtp,rankin.caqg,rankin.9404023}.

Among other results,\footnote{Full details are available in section \ref{discussquat} by simply retricting the results there to be real, reversing the sign used there for 
$ j^2 $, and setting 
$ \tau = 1 $.} this action will give that 
$ \hat \beta = 1 - [ ( 4 \sigma ) / C ] $, and that the electromagnetic four current is proportional to 
$ \hat v^{\mu} $, thus giving 
$ \hat v^{\mu}_{\; \; \| \mu} = 0 $ from conservation of charge. In the original Weyl case, 
$ \hat \beta = 0 $, and there is no contribution to the stress tensor from the constraint terms. Because 
$ \hat v_{\mu} = v_{\mu} - [ ( 1 / 2 ) \ln{ B } ]_{, \mu} $ here, if 
$ C - \hat R \approx C $, the constraint itself clearly takes the form of the Hamilton-Jacobi equation for a nonzero rest mass charged particle in the combined gravitational and electromagnetic field\cite{rankin.ijtp}. The vanishing of the covariant divergence of the stress tensor required by Einstein's equations (using the 
``$ {}_{\|} $'' derivative), also gives results consistent with this interpretation, but with a twist noted below.

In order to make sense of the large dimensionless cosmological constant of 
$ 1 / 4 $, it appears the scale factor 
$ b_{0} = \Lambda / \sigma $ where 
$ \Lambda $ is the usual cosmological constant in laboratory units. Thus, it seems 
$ b_{0} $ must be tiny to make sense of the original Weyl action,\footnote{Actually, Weyl's original treatment\cite{weyl.stm,adler.london} uses coordinates that still have dimensions when taking derivatives, introducing possible confusion on scales.} although a much smaller value for 
$ \sigma $ would seem to relieve that difficulty, while only deviating slightly from Weyl's original action\cite{rankin.9404023}. However, 
$ \sigma $ is eventually found to be irrelevant for this action, as will be noted in comments on 
$ \sigma $ at the end of this section. The bigger factor here is that if the charge in the coupling constant is electronic, the 
$ 6 \hat v^{\mu} \hat v_{\mu} $ in the action with no gravitational constant in front of it will then restrict the value of 
$ b_{0} $ to be near Planck scale values. That assumes 
$ 6 \hat v^{\mu} \hat v_{\mu} $ should be the same order of magnitude as the 
$ j^{2} ( \hat y_{\mu \nu} \hat y^{\mu \nu} ) $ term in the action, where
\begin{equation}
j^{2} = \left ( 8 / g^{2} \right ) G 
\left ( b_{0} / c^{4} \right )
\label{def.j2}
\end{equation}
Here, 
$ G $ is Newton's constant, and for standard electromagnetism coupled through an electronic charge, 
$ g = ( 2 e ) / ( \hbar c ) $ in Gaussian units, and 
$ A_{\mu} = ( \Phi , - \vec A ) $\cite{rankin.1101.3606,jackson}. This seems to be a more likely assumption for the value of 
$ b_{0} $. 

As noted, one interesting consequence of this action is that an element of the electromagnetic sources (which are proportional to 
$ \hat v_{\mu} $) moves like a particle whose charge is fixed by the coupling constant absorbed into 
$ v_{\mu} $ in equation (\ref{guidry.pot}), and whose rest mass is that of the rest mass term in the Hamilton-Jacobi equation in the constraint. That rest mass term is near Planck scale rest mass for near Planck scale 
$ b_{0} $, {\it unless} the 
$ \hat R $ term is large enough to change that. Evidently, the sign of the charge is effectively reversed by reversing the sign of the energy (four vector) in the Hamilton-Jacobi equation. In fact, the normal Lorentz force law\cite{jackson} seems to correspond to negative energy solutions, oddly enough. Otherwise a check shows the sign would be wrong for normal Lorentz force when using the result from the vanishing of the covariant divergence of the stress tensor mentioned above.

However, this action is dependent upon the term 
$ 6 \hat v^{\mu} \hat v_{\mu} $ outside the constraint to achieve those neat results,\footnote{The terms 
$ \hat R + 6 \hat v^{\mu} \hat v_{\mu} $ are essentially just 
$ \hat B_{\mu \nu} \hat g^{\mu \nu} $, ignoring a noncontributing, divergence term.} and that term is unusual. If it is omitted (or if 
$ 16 $ times the contraction of the torsion with itself, is subtracted from the action density), then that is equivalent to injecting the negative of this term times 
$ B^{2} $ into the original Weyl action, and expanding that into the gauge dependent Weyl variables. That's then a higher order action, and it generally produces (via the constraint term contributions in the gauge invariant variables action) both positive energy density fields, plus additional, ghost, negative energy density fields\footnote{My thanks to Jim Wheeler for making these observations to me privately.} with identical equations of motion to those of the positive energy density fields,\footnote{There are subtle differences in higher order terms however, once 
$ \hat R $ is replaced with its value.} but with opposite signed contributions to both the stress tensor and the current four vector\cite{rankin.9404023}. On the one hand, the singularity theorems that afflict classical General Relativity\cite{wald} then no longer need apply, but on the other hand, it may be an understatement to say that this is uncommon\cite{l.h.ford}, even if some papers are exploring negative energy density fields\cite{pos.neg.pair}. A perhaps more intriguing result is that the constraint now takes on the form of the Klein-Gordon equation from wave mechanics in an electromagnetic and gravitational field\cite{rankin.ijtp,galehouse.ijtp.1,galehouse.ijtp.2,rankin.caqg,rankin.9404023} (by substituting 
$ B = \psi^{-2} $), with the standard added conformal term\cite{wald} and a nonzero rest mass, rather than a Hamilton-Jacobi type form. Correctly formed wave function stress tensor and four current terms also appear, and appear consistent with atomic scale Klein-Gordon equation physics if 
$ b_{0} $ is set using atomic instead of Planck level scales. Furthermore with an added self dual, antisymmetric part to the metric, the second order Dirac equation form can be seen, and in quaternions, there are enough correct degrees of freedom to correspond to a two row, complex spinor\cite{rankin.1101.3606,rankin.caqg,rankin.9404023}. Additionally, in a little known paper, Soviet physicist Yuri Usachev shows that the Dirac spinor can also be successfully treated as a scalar (which could be consistent with a quaternionic scalar), and yet still represent spin 
$ 1 / 2 $ fermions correctly\cite{jtep.usachev}. These points will be elaborated further below as part of the results from a fully quaternionic action.

Before continuing to that quaternionic action, more could be said here about the above model resulting from Weyl's original action principle. For example, Maxwell's equations in 
$ \hat v_{\mu} $ seem to be a modified Proca equation with nonlinear modifications introduced from the Ricci tensor. Additionally, the 
$ C $ term in the constraint effectively renders the value of 
$ \sigma $ almost irrelevant because it allows 
$ \sigma $ to be eliminated from the gravitational equations with a little effort. Moreover, because some of the terms in the stress tensor have no gravitational constant as a factor in their value, it's not clear if the approximation 
$ C - \hat R \approx C $ is really valid in many cases, although if 
$ C - \hat R $ is greater than zero and approximately constant, that would not significantly modify the above points. However, the primary purpose of this entire example has been to provide a simple introductory illustration of some of the basic features of actions formed from the gauge invariant variables, not to do a detailed examination of this limited, although historically important particular case. Therefore, that case will not be detailed further here.

\section{Comment on Torsion and Translations}
\label{torsion}

The presence of gauge potentials in torsion has been criticized on general grounds as being contrary to the geometric relation between torsion and translations, and thus to energy-momentum\cite{gronwald.hehl}. The model above achieves its generalization of Weyl's geometry into the quaternions in a neat manner by necessarily shifting some of Weyl's nonmetricity into torsion, and so appears to conflict with this general criticism. This requires an additional comment.

As Einstein noted, torsion is not invariant under his ``lambda transformation''\cite{einstein.meaning}, which is essentially a treatment of the affine connection as a gauge potential rather than as the gauge field that would be the case in equation (\ref{gauge.gamma.weyl}). This difference is crucial to generalizing Weyl's structure into quaternions. Thus it seems odd that a gauge potential could not be part of the torsion to correspond to this fact.

Fortunately, a straightforward resolution of this conflict is already suggested in section \ref{discuss} using the simple real variable action given in equation (\ref{orig.gi.weyl.action}), and its resulting dynamics. As noted there, under simple assumptions, the constraint then becomes the mathematical Hamilton-Jacobi equation for a (nonzero rest mass) charged particle in the combined electromagnetic and gravitational fields. A moment's reflection then reveals that the gauge invariant potential 
$ \hat v_{\mu} $ is essentially the mechanical portion of the four momentum of the ``particle''\cite{goldstein}. Thus the {\it gauge invariant torsion} will be a three index tensor antisymmetric in the two covariant indices, directly constructed from that four momentum, and the Kronecker delta, times a constant. Then the torsion is indeed related to energy-momentum, and that energy-momentum also appears as a source of the gravitational field through the stress tensor. This would seem to resolve the conflict in a novel manner, one which actually reaffirms the criteria specified by Hehl and Gronwald in their criticism\cite{gronwald.hehl}. The spirit of that observation survives when other actions are chosen\cite{rankin.caqg}, such as in the next section, so it seems likely to be a more general property of this kinematical structure.

\section{A Quaternionic Action}
\label{discussquat}

Weyl's original, real action provides a quick survey of the mathematics of this structure with its use of gauge invariant variables and their associated constraint in the action. However, a much more general, quaternionic action is needed to discover how a fully quaternionic structure affects that mathematics. Now the action of equation (\ref{orig.gi.weyl.action}) was not just a simple extension of a standard action for gravitation and electromagnetism\cite{rankin.caqg} because it contained an additional 
$ 6 \hat v^{\mu} \hat v_{\mu} $ term. As noted then, if 
$ 16 $ times the contraction of the torsion with itself is subtracted from that action density, then those extra terms are removed. However, if instead 
$ 16 \kappa $ times the contraction of the torsion with itself is subtracted, then the action contains an additional term 
$ 6 \tau \hat v^{\mu} \hat v_{\mu} $ where 
$ \tau = 1 - \kappa $, and 
$ \kappa $ and 
$ \tau $ are real constants. Using these parameters, both the original Weyl action form, and the simpler standard Einstein and Yang-Mills action are special cases of the results, for 
$ \tau = 1 $ and 
$ \tau = 0 $ respectively. Thus this 
$ \tau $ parametrized action will now be used, with the additional step of including the full, quaternionic Yang-Mills field.

Since it was already noted in section \ref{discuss} that the 
$ \tau = 0 $ case for the action includes negative energy density fields for solutions in the complex plane\cite{rankin.9404023}, this case may seem to be an odd choice to include. However, even the 
($ \tau = 1 $) generalized, original, real Weyl action (equation (\ref{orig.gi.weyl.action})) ultimately produces a term 
$ ( C / 4 ) \hat g_{\mu \nu} $ for the cosmological constant term in the gravitational equation, because 
$ \hat \beta = 1 - [ ( 4 \sigma ) / C ] = 0 $. By substituting for 
$ \hat v^{\mu} \hat v_{\mu} $ from the constraint, an additional 
$ ( C / 2 ) \hat g_{\mu \nu} $ also will appear (from the 
$ 3 \tau \hat v_{0}^{\mu} \hat v_{0 \mu} \hat g^{\mu \nu} $ term in equation (\ref{quat.einstein.result})). This might be interpreted either as part of the cosmological constant term, or as a ``rest mass'' related stress-energy tensor term. Either way, clearly one sign choice for 
$ C $ will produce a negative overall energy density, even in that simple model. And, a no-frills, quaternionic action form without unfamiliar extra terms beyond the constraint terms should be instructive, even if ultimately exotic in some respects. Furthermore, the 
$ \tau = 1 $ original Weyl choice for the action will be seen to produce results that are no longer as simple when extended into quaternions, and it will even produce additional negative energy densities in terms {\it not} associated with the constraint. Thus that case will lose some of its appealing simplicity.

Therefore, adopt the quaternionic action
{\samepage 
\begin{eqnarray}
I & = & \int { \sqrt{-\hat g}\, \left \{ 
\left ( \hat R - 2 \sigma \right ) + 
\frac{1}{2} \left [ \frac{1}{2} j^{2} \left ( 
\hat y^{\mu \nu} \hat y_{\mu \nu} \right ) 
\right. \right. } 
\nonumber \\
 &   & + \left. \left. \left ( \hat R + 6 
\hat v^{\mu}_{\; \; \| \mu} + 6 \hat v^{\mu} 
\hat v_{\mu} - C \right ) \hat \beta \right. \right. 
\nonumber \\
 &   & + \left. \left. 6 \tau \hat v^{\mu} 
\hat v_{\mu} + QC \right ] \right \} d^{4}x 
\label{quat.weyl.action}
\end{eqnarray}}%
where the 
$ QC $ term is the quaternion conjugate of everything preceding it in that level of brackets. The reversal in the sign of the gravitational constant 
$ j^{2} $ will be explained shortly.

The variation of this action is most simply carried out by varying the individual real components of each of the quaternionic quantities involved. To facilitate that, adopt a component notation in which any quaternionic quantity 
$ Z $ is written out in its real components as
\begin{equation}
Z = Q_{0} Z_{0} + \sum_{ k = 1 }^{3} Q_{k} Z_{k} = 
Z_{0} + \vec Q {\bf \cdot} \vec Z
\label{quat.components}
\end{equation}
where the 
$ Z_{a} $ are all real, and 
$ a $ is the quaternion component index, with 
$ a $ ranging from 
$ 0 $ to 
$ 3 $. The matching vector notation should be self explanatory. Any additional lower indices such as tensor indices, will physically follow the quaternion component index if lower tensor indices are present for the quantity in question.

In terms of those real components of quaternionic quantities, equation (\ref{quat.weyl.action}) now becomes
{\samepage 
\begin{eqnarray}
I & = & \int { \sqrt{-\hat g}\, \left \{ 
\left ( \hat R - 2 \sigma \right ) + 
\frac{1}{2} j^{2} \left ( 
\hat y_{0}^{\mu \nu} \hat y_{0 \mu \nu} \right. 
\right. } 
\nonumber \\
 &   & \left. \left. - {\vec {\hat y}}{}^{\mu \nu} 
{\bf \cdot} {\vec {\hat y}}_{\mu \nu} \right ) + 
\left ( \hat R + 6 \hat v_{0 \| \mu}^{\mu} + 
6 \hat v_{0}^{\mu} \hat v_{0 \mu} - 6 {\vec {\hat 
v}}{}^{\mu} {\bf \cdot} {\vec {\hat v}}_{\mu} \right. 
\right. 
\nonumber \\
 &   & \left. \left. - C \right ) \hat \beta_{0} 
- 6 \left ( {\vec {\hat v}}{}_{\; \; \| \mu}^{\mu} 
+ 2 \hat v_{0}^{\mu} {\vec {\hat v}}_{\mu} \right ) 
{\bf \cdot} {\vec {\hat \beta}} \right. 
\nonumber \\
 &   & + \left. 6 \tau \hat v_{0}^{\mu} 
\hat v_{0 \mu} - 6 \tau {\vec {\hat v}}{}^{\mu} 
{\bf \cdot} {\vec {\hat v}}_{\mu} \right \} d^4x 
\label{quat.comp.weyl.action}
\end{eqnarray}}%
where the added quaternion conjugate terms have already been explicitly included in this result. This action is used in conjunction with the component form of equation (\ref{def.yhat})
\begin{equation}
\hat y_{a \mu \nu} = \hat v_{a \nu , \mu} - 
\hat v_{a \mu , \nu} + 4 \sum_{ k = 1 }^{3} 
\sum_{ n = 1 }^{3} \epsilon_{0 a k n} 
\hat v_{k \nu} \hat v_{n \mu}
\label{def.comp.yhat}
\end{equation}
where lowercase Latin indices range from 
$ 0 $ to $ 3 $ unless otherwise noted, and 
$ \epsilon_{a b c d} $ is the totally antisymmetric unit symbol with 
$ \epsilon_{0 1 2 3} = 1$. This action is a functional of 
$ \hat g^{\mu \nu} $, 
$ \hat v_{a \mu} $, and 
$ \hat \beta_{a} $.

The variation of the components 
$ \hat \beta_{a} $ yields four equations which reassemble into the quaternionic form of the required constraint of equation (\ref{def.identity.q}). Just as with the real case in section \ref{discuss}, this equation in principle gives 
$ B $ when 
$ \hat g_{\mu \nu} $ and 
$ v_{\mu} $ are already known, the latter being obtained in some arbitrary gauge. The solution for 
$ B $ then gives 
$ \hat v_{\mu} $ from 
$ v_{\mu} $ by equation (\ref{def.qvhat}). However, this can be a challenging problem to solve in the full quaternions compared to cases in the complex plane\cite{rankin.caqg,rankin.9404023,rankin.1101.3606}. Nevertheless, in the free space case of Lorentzian spacetime and vanishing Yang-Mills field, the constraint has simple solutions constructed from solutions to linear wave equations, including the standard solutions for a free Dirac particle. This will be shown shortly.

\subsection{Equations of Motion}
\label{quat.eq.motion}

The variation of the components 
$ \hat v_{a \mu} $ yields four, four-vector equations, which reassemble into the quaternionic form of the Yang-Mills field equations. That gives the deceptively simple looking equation
{\samepage 
\begin{eqnarray}
\hat y_{\; \; \| \nu}^{\mu \nu} + 2  \left ( 
\hat y^{\mu \nu} \hat v_{\nu} - \hat v_{\nu} 
\hat y^{\mu \nu} \right ) = \frac{3}{j^{2}} 
\left [ \hat g^{\mu \nu} \hat \beta_{, \nu} \right. 
\nonumber \\
 - \left. \left ( \hat v^{\mu} \hat \beta 
+ \hat \beta \hat v^{\mu} \right ) - 2 \tau 
\hat v^{\mu} \right ]
\label{yang.mills.quat}
\end{eqnarray}}%
with the second and third terms on the left or field side forming a commutator, and the second and third terms on the right or ``source'' side forming an anticommutator.\footnote{As a side observation, equations (\ref{def.qvhat}) and (\ref{def.yhat}) give that 
$ \hat y_{\; \; \| \nu}^{\mu \nu} + 2  ( \hat y^{\mu \nu} \hat v_{\nu} - 
\hat v_{\nu} \hat y^{\mu \nu} ) = B^{-1} [ y_{\; \; \| \nu}^{\mu \nu} 
+ 2  ( y^{\mu \nu} v_{\nu} - v_{\nu} y^{\mu \nu} ) ] B $.} When everything is real in the 
$ \tau = 0 $ case, and the sign reversal adopted for 
$ j^{2} $ in equation (\ref{quat.weyl.action}) is remembered, this reduces exactly to equation (3.8) in an earlier paper\cite{rankin.caqg}.

The variation of the components 
$ \hat g^{\mu \nu} $ yields the gravitational Einstein equations. For these, quaternions will remain stated in components, and that gives
{\samepage 
\begin{eqnarray}
\hat R_{\mu \nu} - \frac{1}{2} \hat R \hat g_{\mu \nu} + 
\sigma \hat g_{\mu \nu} & = & j^2 \left [ \left ( 
\hat y_{0 \mu}^{\; \; \; \alpha} 
\hat y_{0 \alpha \nu} + \frac{1}{4} \hat g_{\mu \nu} 
\hat y_{0}^{\alpha \rho} \hat y_{0 \alpha \rho} \right ) \right. 
\nonumber \\
 &   & - \left. \left ( {\vec {\hat y}}{}_{\mu}^{\; \; \alpha} 
{\bf \cdot} {\vec {\hat y}}_{\alpha \nu} + \frac{1}{4} 
\hat g_{\mu \nu} {\vec {\hat y}}{}^{\alpha \rho } {\bf \cdot} 
{\vec {\hat y}}_{\alpha \rho } \right ) \right ] 
\nonumber \\
 &   & - \left ( \hat R_{\mu \nu} - \frac{1}{2} 
\hat R \hat g_{\mu \nu} \right ) \hat \beta_{0} - 
\hat \beta_{0 \| \mu \| \nu} 
\nonumber \\
 &   & + \hat \beta_{0 \| \alpha \| \rho} \hat g^{\alpha \rho} 
\hat g_{\mu \nu} + 3 \left ( \hat \beta_{0 , \mu} 
\hat v_{0 \nu} \right. 
\nonumber \\
 &   & + \left. \hat \beta_{0 , \nu } \hat v_{0 \mu} 
\right ) - 3 \hat \beta_{0 , \alpha} \hat v_{0}^{\alpha} 
\hat g_{\mu \nu} 
\nonumber \\
 &   & - 6 \hat v_{0 \mu} \hat v_{0 \nu} \hat \beta_{0} 
+ 3 \hat v_{0}^{\alpha} \hat v_{0 \alpha} 
\hat g_{\mu \nu} \hat \beta_{0} 
\nonumber \\
&   & + 6 {\vec {\hat v}}_{\mu} {\bf \cdot} {\vec {\hat v}}_{\nu} 
\hat \beta_{0} - 3 {\vec {\hat v}}{}^{\alpha} {\bf \cdot} 
{\vec {\hat v}}_{\alpha} \hat g_{\mu \nu} \hat \beta_{0} 
\nonumber \\
 &   & - \frac{1}{2} C \hat g_{\mu \nu } \hat \beta_{0} - 
3 \left ( {\vec {\hat \beta}}_{, \mu} {\bf \cdot} 
{\vec {\hat v}}_{\nu} \right. 
\nonumber \\
 &   & \left. + {\vec {\hat \beta}}_{, \nu} {\bf \cdot} 
{\vec {\hat v}}_{\mu} \right ) + 3 {\vec {\hat \beta}}_{, \alpha} 
{\bf \cdot} {\vec {\hat v}}{}^{\alpha} \hat g_{\mu \nu} 
\nonumber \\
 &   & + 6 \left ( \hat v_{0 \mu} {\vec {\hat v}}_{\nu} + 
\hat v_{0 \nu} {\vec {\hat v}}_{\mu} \right ) {\bf \cdot} 
{\vec {\hat \beta}} 
\nonumber \\
 &   & - 6 \hat g_{\mu \nu} \hat v_{0}^{\alpha} 
{\vec {\hat v}}_{\alpha} {\bf \cdot} {\vec {\hat \beta}} - 
3 \tau \left [ 2 \hat v_{0 \mu} \hat v_{0 \nu} \right. 
\nonumber \\
 &   & - \hat v_{0}^{\alpha} \hat v_{0 \alpha} 
\hat g_{\mu \nu} - 2 {\vec {\hat v}}_{\mu} {\bf \cdot} 
{\vec {\hat v}}_{\nu} 
\nonumber \\
 &   & + \left. {\vec {\hat v}}{}^{\alpha} {\bf \cdot} 
{\vec {\hat v}}_{\alpha} \hat g_{\mu \nu} \right ]
\label{quat.einstein.result}
\end{eqnarray}}%
This immediately illustrates why 
$ j^{2} $ has been given a reversed sign in the action. Clearly the Yang-Mills field stress tensor term containing 
$ \hat y_{0 \mu \nu} $ has a sign opposite to the stress tensor term for 
$ {\vec {\hat y}}_{\mu \nu} $ contributed by the purely imaginary part of the Yang-Mills field.\footnote{Given quaternions 
$ Z $ and 
$ \hat Z $ such that 
$ \hat Z = B^{-1} Z B $, where 
$ Z $ may have tensor indices, then 
$ ( \hat Z + QC ) = B^{-1} ( Z + QC ) B = ( Z + QC ) $. Then 
$ \hat y_{a \mu \nu} \rightarrow y_{a \mu \nu} $ is allowed in the 
$ j^{2} $ stress-energy tensor terms.} Thus, the sign of 
$ j^2 $ has been chosen so that the totally imaginary portion of the field has a stress tensor term which behaves like a ``normal'' or positive energy density, while the real portion produces negative energy density. Since standard Yang-Mills theory works with the totally imaginary subset of all quaternion valued Yang-Mills fields, those fields have been assigned positive energy density.\footnote{The real part 
$ \hat y_{0 \mu \nu} $ could be constrained to vanish so that the reversed sign energy density term is absent from the outset. The term 
$ \hat \alpha^{\mu \nu} \hat y_{0 \mu \nu} $ is simply added to the action density, and the real, antisymmetric quantity 
$ \hat \alpha^{\mu \nu} $ is also varied in the action.}

However, for 
$ \tau \neq 0 $, a similar situation also arises with some reversed sign energy density terms in the 
$ \tau $ terms in the stress tensor, and if the terms for the imaginary part of 
$ \hat v_{\mu} $ are dominant, only 
$ \tau \leq 0 $ would suppress the reversed sign energy density which arises in those terms. But 
$ \tau \leq 0 $ does {\it not} include the generalization of the original Weyl action 
($ \tau = 1 $) into the complex and quaternion numbers. Thus, the generalization of the original Weyl action into the quaternions already appears to allow negative energy density from the 
$ \tau \neq 0 $ terms in the stress tensor, even without considering the stress tensor terms arising from the constraint.

Now two more equations can be derived immediately from equations (\ref{def.yhat}), (\ref{def.identity.q}), (\ref{yang.mills.quat}), and (\ref{quat.einstein.result}), together with
\begin{equation}
\hat y_{\; \; \| \nu \| \mu}^{\mu \nu} = 
\frac{1}{\sqrt{ - \hat g }} \left ( \sqrt{ - \hat g } 
\, \hat y^{\mu \nu} \right )_{, \nu , \mu} = 0
\label{div.div.yhat}
\end{equation}
This last equation combines with the 
``$ {}_{\| \mu} $'' divergence of equation (\ref{yang.mills.quat}), together with equation (\ref{def.yhat}), and equation (\ref{yang.mills.quat}), to give that
{\samepage 
\begin{eqnarray}
\hat g^{\mu \nu} \hat \beta_{\| \mu \| \nu} & - & 
\hat v^{\mu}_{\; \; \| \mu} \hat \beta - 
\hat \beta \hat v^{\mu}_{\; \; \| \mu} - 3 \hat v^{\mu} 
\hat \beta_{, \mu} 
\nonumber \\
 & + & \hat \beta_{, \mu} \hat v^{\mu} - 2 \hat \beta 
\hat v^{\mu} \hat v_{\mu} + 2 \hat v^{\mu} 
\hat v_{\mu} \hat \beta = 2 \tau \hat v^{\mu}_{\; \; \| \mu}
\label{consv.charge}
\end{eqnarray}}%
This can be considered the conservation equation for quaternionic charge, and is the first of the two additional derived equations mentioned above. Notice that for 
$ \hat \beta $ equal to a real constant other than 
$ - \tau $, it gives 
$ \hat v^{\mu}_{\; \; \| \mu} = 0 $, while a nonconstant 
$ \hat \beta $ generally gives {\it no} such simple result.

Using equation (\ref{def.identity.q}) judiciously on this equation to substitute for some terms or portions of terms (a term may be split into the sum of two equal halves), allows this equation eventually to be rewritten as
{\samepage 
\begin{eqnarray}
\frac{3}{2} \left [ \hat g^{\mu \nu} \hat \beta_{\| \mu \| \nu} - 
\hat v^{\mu}_{\; \; \| \mu} \hat \beta - 2 \hat v^{\mu} 
\hat \beta_{, \mu} \right. 
\nonumber \\
\left. + \hat v^{\mu} \hat v_{\mu} \hat \beta - \frac{1}{6} 
\left ( C - \hat R \right ) \hat \beta \right ] 
\nonumber \\
 - \frac{1}{2} \left [ \hat \beta_{\| \mu \| \nu} 
\hat g^{\mu \nu} - \hat \beta \hat v^{\mu}_{\; \; \| \mu} - 
2 \hat \beta_{, \mu} \hat v^{\mu} \right. 
\nonumber \\
\left. + \hat \beta \hat v^{\mu} \hat v_{\mu} - \frac{1}{6} 
\hat \beta \left ( C - \hat R \right ) \right ] = 
2 \tau \hat v^{\mu}_{\; \; \| \mu}
\label{consv.charge.kg}
\end{eqnarray}}%
The left side is an asymmetric combination of a right handed Klein-Gordon equation form (if the square brackets vanish) and a left handed Klein-Gordon equation form, with 
$ - \hat v_{\mu} $ in place of 
$ v_{\mu} $, 
$ \hat \beta $ in place of the wavefunction, and with the addition of the standard (real) 
$ ( 1 / 6 ) \hat R $ ``conformal'' term coefficient\cite{rankin.9404023} times 
$ \hat \beta $ in each case. The 
$ - ( 1 / 6 ) C \hat \beta $ term acts as the ``rest mass'' squared term, and has the proper sign if 
$ C < 0 $. If 
$ \hat \beta $ commutes with 
$ \hat v_{\mu} $ and its derivatives, such as when everything is in a complex plane, or when 
$ \hat \beta $ is real, then the two parts merge into a single, Klein-Gordon equation form with the reversed sign, gauge invariant potential, although for nonzero 
$ \tau $, it has an additional, inhomogenous term in that equation. This result becomes the equation of motion of 
$ \hat \beta $, and immediately illustrates why negative energy densities might arise from the constraint terms in this set of equations. Particularly for 
$ \tau = 0 $, equation (\ref{consv.charge.kg}) appears to be linear in 
$ \hat \beta $, while 
$ \hat \beta $ appears linearly in its stress tensor terms in equation (\ref{quat.einstein.result}) (and the quaternionic ``four-current sources'' in equation (\ref{yang.mills.quat})), rather than quadratically as a square of either a real number or an absolute value of a quaternion.\footnote{Note that 
$ \hat \beta = 0 $ is always a possible solution to equation (\ref{consv.charge.kg}) if 
$ \tau = 0 $.} That means sign reversals in the stress tensor terms 
may occur.

The other additional equation results from the trace of equation (\ref{quat.einstein.result}). Combine that with the real part of equation (\ref{consv.charge}), and use equation (\ref{def.identity.q}) as well as the real part of a product of that equation with 
$ \hat \beta $. That will simplify the trace equation to
\begin{equation}
\left ( 1 - \tau \right ) \hat R - 4 \sigma = 
C \left ( \hat \beta_{0} - \tau \right )
\label{rhat}
\end{equation}
For most values of 
$ \tau $, including zero, this relates 
$ \hat R $ and 
$ \hat \beta_{0} $. However for the special value 
$ \tau = 1 $, this can be solved instead for 
$ \hat \beta_{0} = 1 - 4 ( \sigma / C ) $, which is constant. However even then, only 
$ \hat \beta_{0} $ is set, while the three imaginary components of 
$ \hat \beta $ are not determined through this equation. Even in the complex plane, the one imaginary component is not set. Outside the pure real numbers, 
$ \hat \beta $ is not necessarily constant for 
$ \tau = 1 $, and thus generally 
$ \hat v^{\mu}_{\; \; \| \mu} \neq 0 $ even for 
$ \tau = 1 $. That suggests the 
$ \tau = 1 $ case is no longer so simple in the complex and quaternion numbers, unlike the case of real numbers in section \ref{discuss}. Since as already noted, this 
$ \tau = 1 $ case may allow negative energy density in the 
$ \tau $ terms in the stress tensor, the 
$ \tau = 0 $ (Einstein / Yang-Mills) case may be simpler and no more exotic in nature overall than the 
$ \tau = 1 $ case now. Thus, 
$ \tau = 0 $ will be the case assumed hereafter unless otherwise noted.

Since 
$ \tau = 0 $, notice that none of the non-field terms (terms not containing 
$ \hat y_{\mu \nu} $) on the right hand side of equation (\ref{quat.einstein.result}) involve a gravitational constant, but they all {\it do} include 
$ \hat \beta $. Its magnitude will serve in place of a gravitational constant,\footnote{Attempts to relate 
$ \hat \beta $ to actual physics would be expected to incorporate Newton's constant 
$ G $ into the magnitude of 
$ \hat \beta $ explicitly at some point\cite{rankin.1101.3606}.} and equation (\ref{rhat}) directly reflects that. The action of equation (\ref{quat.weyl.action}) no longer shares the problem of the action of equation (\ref{orig.gi.weyl.action}) in which terms appear with no moderating effects of either a gravitational constant, or 
$ \hat \beta $. Thus the current action avoids the problem of section \ref{discuss} in which 
$ b_{0} $ had to be set such that 
$ j^{2} $ (defined by equation (\ref{def.j2})) was of order unity. That was necessary there in order to obtain consistency of stress tensor term magnitudes. For the current action, 
$ b_{0} $ (and thus 
$ j^{2} $) is still free at this point. Furthermore once the term 
$ - \left ( \hat R_{\mu \nu} - \frac{1}{2} \hat R \hat g_{\mu \nu} \right ) 
\hat \beta_{0} $ is moved from the right side of equation (\ref{quat.einstein.result}) to the left side of that equation, it becomes clear that eventually the quantity 
$ ( 1 + \hat \beta_{0} ) $ will wind up in the denominator of the right side of Einstein's equation, as well as the cosmological constant term. Then in regions in which 
$ \hat \beta_{0} \rightarrow - 1 $, the gravitational coupling constant 
$ j^{2} $ and cosmological constant 
$ \sigma $ might be (incorrectly) interpreted as ``running'' to ever larger magnitudes, and the same might apply to the other ``source'' terms remaining on the right side of Einstein's equation. Note that a negative 
$ \hat \beta_{0} $ is ``normal'' positive energy density in the term 
$ ( 1 / 2 ) C \hat g_{\mu \nu} \hat \beta_{0} $ in equation (\ref{quat.einstein.result}) if 
$ C $ is negative, so this case need not be exotic. 
However, this analysis so far applies only to the coupling of the gravitational field to the stress-energy tensor, and does not include the running of any other coupling constants, such as is encountered in standard quantum field theory\cite{guidry}.\footnote{On the other hand, this result {\it does} demonstrate the type of relatively simple mathematics that will produce apparent ``running'' of any coupling constant. An electromagnetic coupling example in the complex plane appears in earlier work\cite{rankin.caqg}.}

\subsection{Interpreting the Constraint and Its Effects}
\label{examine.constraint}

As expected, the action of equation (\ref{quat.weyl.action}) has produced the familiar Einstein and Yang-Mills equations, with an additional real component in the Yang-Mills field, and the entire Yang-Mills field appearing in the stress-energy tensor in Einstein's equation. However, the added constraint terms in the action also produce possible ``sources'' in both field equations, and introduce the constraint itself as another equation to solve. In that regard, equation (\ref{consv.charge.kg}) makes it clear that one solution for the Lagrange multiplier 
$ \hat \beta $ is 
$ \hat \beta = 0 $, and that illustrates immediately that all those additional ``source'' terms may vanish. That produces an Einstein-Yang-Mills equation system in which only the self interaction of the Yang-Mills field generated by the term 
$ 2  ( \hat y^{\mu \nu} \hat v_{\nu} - \hat v_{\nu} \hat y^{\mu \nu} ) $ in equation (\ref{yang.mills.quat}) acts as a source for the Yang-Mills field. The constraint equation (\ref{def.identity.q}) in principal gives 
$ \hat v_{\mu} $ by solving for 
$ B $, given 
$ \hat g_{\mu \nu} $, and 
$ v_{\mu} $ in any arbitrary gauge. When working with fully quaternionic quantities, this is nontrivial. This subsection will analyze a very simple case which can be reasonably well handled analytically, the free case in which 
$ v_{\mu} $ as well as 
$ y_{\mu \nu} $ vanish in some gauge. Additionally, the metric 
$ \hat g_{\mu \nu} $ will be assumed to be Lorentzian with signature 
$ ( + - - - ) $. If the analog between the constraint and mechanics noted in section \ref{discuss} is maintained, this could be called the ``free particle'' case.

However, even if that analogy is inappropriate, the constraint and its associated terms in the equations of motion remain an integral part of the mathematical structure of the gauge invariant variables used to describe the system, and that will be true of all actions expressed in those variables. That means the constraint and its results must still be investigated, and an effort made to interpret the results as part of the predicted physics for any such actions, including the action of equation (\ref{quat.weyl.action}). This is an automatic part of the use of these gauge invariant variables. Obviously, cases that produce 
$ \hat \beta = 0 $ will minimize, but still not entirely remove this concern, as the action of section \ref{discuss} illustrates when Weyl's original parameters are used. Any further insights require additional examples, including the one generated by the action of equation (\ref{quat.weyl.action}) (assuming 
$ \tau = 0 $) that will now be examined.

As a preliminary, note that equations (\ref{def.identity.q}) and (\ref{def.qvhat}) can be used to show that
\begin{equation}
B \left ( \hat v^{\mu}_{\; \; \| \mu} + \hat v^{\mu} 
\hat v_{\mu} \right ) B^{-1} = 
\frac{1}{6} \left ( C - \hat R \right )
\label{b.vs.term.binverse}
\end{equation}
and thus
{\samepage 
\begin{eqnarray}
\frac{1}{6} \left ( C - \hat R \right ) & = & 
 - \frac{3}{2} \hat g^{\mu \nu} B_{, \mu} B^{-1} 
v_{\nu} + \frac{1}{2} \hat g^{\mu \nu} v_{\nu} 
B_{, \mu} B^{-1} 
\nonumber \\
 &   & + \frac{1}{\sqrt{ - \hat g }} \left ( \sqrt{ - \hat g } 
\, \hat g^{\mu \nu} v_{\nu} \right )_{, \mu} 
+ \hat g^{\mu \nu} v_{\mu} v_{\nu} 
\nonumber \\
 &   & + \frac{3}{4} \hat g^{\mu \nu} B_{, \mu} B^{-1} 
B_{, \nu} B^{-1} 
\nonumber \\
 &   & - \frac{1}{2} \frac{1}{\sqrt{ - \hat g }} \left ( 
\sqrt{ - \hat g } \, \hat g^{\mu \nu} B_{, \nu} 
\right )_{, \mu} B^{-1} 
\label{quat.expanded.constraint}
\end{eqnarray}}%
Because of the quaternionic nature of this equation, the substitution 
$ B = \psi^{-2} $ will not linearize this equation as it does in the complex plane\cite{rankin.caqg,rankin.9404023}. Therefore, assume
\begin{equation}
B = \chi^{-1} \psi^{-1}
\label{linear.change.quat}
\end{equation}
where
\begin{equation}
\chi_{, \mu} \chi^{-1} = \psi^{-1} \psi_{, \mu}
\label{chi.psi.relate}
\end{equation}
Provided this equation has a solution for 
$ \chi $ and 
$ \psi $ in terms of each other,\footnote{At this point, quaternionic equation (\ref{chi.psi.relate}) will be considered to hold only for a specific gauge for 
$ v_{\mu} $.} this reduces equation (\ref{quat.expanded.constraint}) to
{\samepage 
\begin{eqnarray}
\frac{1}{6} \left ( C - \hat R \right ) & = & 
 3 \hat g^{\mu \nu} \chi^{-1} \chi_{, \mu} 
v_{\nu} - \hat g^{\mu \nu} v_{\nu} 
\chi^{-1} \chi_{, \mu} 
\nonumber \\
 &   & + \frac{1}{\sqrt{ - \hat g }} \left ( \sqrt{ - \hat g } 
\, \hat g^{\mu \nu} v_{\nu} \right )_{, \mu} 
+ \hat g^{\mu \nu} v_{\mu} v_{\nu} 
\nonumber \\
&   & + \frac{ \chi^{-1} }{\sqrt{ - \hat g }} \left ( 
\sqrt{ - \hat g } \, \hat g^{\mu \nu} \chi_{, \nu} 
\right )_{, \mu}
\label{quat.expanded.constraint.chi}
\end{eqnarray}}%
which can be rewritten as
{\samepage 
\begin{eqnarray}
0 & = & \frac{3}{2} \left \{ 
\frac{ \chi^{-1} }{\sqrt{ - \hat g }} 
\left ( \sqrt{ - \hat g } \, \hat g^{\mu \nu} \chi_{, \nu} 
\right )_{, \mu} + \frac{1}{\sqrt{ - \hat g }} \left ( 
\sqrt{ - \hat g } \, \hat g^{\mu \nu} v_{\nu} 
\right )_{, \mu} \right. 
\nonumber \\
&   & + 2 \left. \hat g^{\mu \nu} \chi^{-1} \chi_{, \mu} v_{\nu} 
+ \hat g^{\mu \nu} v_{\mu} v_{\nu} - \frac{1}{6} 
\left ( C - \hat R \right ) \right \} 
\nonumber \\
& - & \frac{1}{2} \left \{ \frac{ \chi^{-1} }{\sqrt{ - \hat g }} 
\left ( \sqrt{ - \hat g } \, \hat g^{\mu \nu} \chi_{, \nu} 
\right )_{, \mu} + \frac{1}{\sqrt{ - \hat g }} \left ( 
\sqrt{ - \hat g } \, \hat g^{\mu \nu} v_{\nu} \right )_{, \mu} 
\right. 
\nonumber \\
&  & + 2 \left. \hat g^{\mu \nu} v_{\nu} \chi^{-1} \chi_{, \mu} 
+ \hat g^{\mu \nu} v_{\mu} v_{\nu} - \frac{1}{6} 
\left ( C - \hat R \right ) \right \} 
\label{quat.constraint.kg}
\end{eqnarray}}%
Obviously if everything commutes, this will reduce to a Klein-Gordon equation (for 
$ C < 0 $). In its current non-commuting quaternionic form, it has somewhat the same peculiar asymmetric combination of forms seen previously in equation (\ref{consv.charge.kg}) if 
$ \tau = 0 $, although the current equation is even less neat in form. However, if 
$ v_{\mu} = 0 $, it does produce the quaternionic Klein-Gordon ``free particle'' equation (with the added conformal term\cite{wald} containing 
$ ( 1 / 6 ) \hat R \, $)
\begin{equation}
\frac{1}{\sqrt{ - \hat g }} \left ( \sqrt{ - \hat g } \, 
\hat g^{\mu \nu} \chi_{, \nu} \right )_{, \mu} = 
\frac{1}{6} \left ( C - \hat R \right ) \chi
\label{quat.free.kg.chi}
\end{equation}
where the metric is assumed to be a flat spacetime metric (which sets 
$ \hat R = 0 $ if that's global). This is the promised simple, linear equation for the ``free particle'' solutions to the constraint equation, provided of course that equation (\ref{chi.psi.relate}) is also solved for 
$ \psi $. If 
$ v_{\mu} = 0 $ is assumed at the outset, then 
$ \hat v_{\mu} = - ( 1 / 2 ) B^{-1} B_{, \mu} $ directly from equation (\ref{def.qvhat}), and then equations (\ref{linear.change.quat}) and (\ref{chi.psi.relate}) give 
$ \hat v_{\mu} = \psi_{, \mu} \psi^{-1} $. Expanding the constraint of equation (\ref{def.identity.q}) from that gives
\begin{equation}
\frac{1}{\sqrt{ - \hat g }} \left ( \sqrt{ - \hat g } \, 
\hat g^{\mu \nu} \psi_{, \nu} \right )_{, \mu} = 
\frac{1}{6} \left ( C - \hat R \right ) \psi
\label{quat.free.kg.psi}
\end{equation}
However, this only indicates that 
$ \psi $ and 
$ \chi $ both obey the Klein-Gordon equation. It does not yet give 
$ \psi $ in terms of 
$ \chi $ by using equation (\ref{chi.psi.relate}).

\subsection{Solving Equation (\ref{chi.psi.relate})}
\label{solve.mirror}

If equation (\ref{chi.psi.relate}) is differentiated using the 
``$ {}_{, \nu} $'' derivative, this gives integrability conditions necessary in order for the equation even to have solutions. Requiring 
$ \chi_{, \mu , \nu} = \chi_{, \nu , \mu} $, and
$ \psi_{, \mu , \nu} = \psi_{, \nu , \mu} $ gives a pair of equations which may be subtracted to give the two integrability conditions by also using equation (\ref{chi.psi.relate}) itself. They are (written as one equation)
\begin{equation}
\left [ \psi^{-1} \psi_{, \mu} , \psi^{-1} \psi_{, \nu} 
\right ] = 0 = 
\left [ \chi_{, \mu} \chi^{-1} , \chi_{, \nu} \chi^{-1} 
\right ]
\label{integrability}
\end{equation}
where the ``$ [ \; , \; ] $'' terms are conventional commutators. These may be simplified to
\begin{equation}
\psi_{, \mu} \psi^{\dagger} \psi_{, \nu} - 
\psi_{, \nu} \psi^{\dagger} \psi_{, \mu} = 0
\label{integrability.simpler.psi}
\end{equation}
\begin{equation}
\chi_{, \mu} \chi^{\dagger} \chi_{, \nu} - 
\chi_{, \nu} \chi^{\dagger} \chi_{, \mu} = 0
\label{integrability.simpler.chi}
\end{equation}
where the 
``$ {}^{\dagger} $'' indicates the quaternion conjugate of the quantity it follows.\footnote{Some of the references use a reversed notation which places the 
``$ {}^{\dagger} $'' and other similar operators to the left of the symbol affected.}

If the quaternion conjugate of equation (\ref{chi.psi.relate}) is taken, and added to the original equation, the result can be written as
\begin{equation}
\frac{ \left ( \chi^{\dagger} \chi \right )_{, \mu} }{ 
\chi^{\dagger} \chi } = 
\frac{ \left ( \psi^{\dagger} \psi \right )_{, \mu} }{ 
\psi^{\dagger} \psi }
\label{chi.psi.amplitudes}
\end{equation}
This states that if equation (\ref{chi.psi.relate}) has a solution, then the squared real norms of 
$ \chi $ and 
$ \psi $ are always proportional. The constant of proportionality may be chosen as one without loss of generality, so that they both have exactly the same real norms, or amplitudes.

Generally, if 
$ \psi_{, \mu} = \lambda_{\mu} \psi $ and 
$ \lambda_{\mu} \lambda_{\nu} = \lambda_{\nu} \lambda_{\mu} $ (``four momentum'' left eigenvalues exist and commute), or if 
$ \psi_{, \mu} = \psi \alpha_{\mu} $ and 
$ \alpha_{\mu} \alpha_{\nu} = \alpha_{\nu} \alpha_{\mu} $ (``four momentum'' right eigenvalues exist and commute), then the integrability conditions specified by equation 
(\ref{integrability.simpler.psi}) are satisfied. Similar conditions on left or right eigenvalues of 
$ \chi_{, \mu} $ will satisfy the integrability conditions of equation (\ref{integrability.simpler.chi}). Furthermore, {\it all} two component or two dimensional cases are easily shown always to satify the integrability conditions. These are cases in which both 
$ \psi $ and 
$ \chi $ can be written in terms of quaternions with only two nonzero components in common directions out of the four possible components in a general quaternion. These then have simple general solutions to equation (\ref{chi.psi.relate}) relating 
$ \chi $ and 
$ \psi $ to each other. In the subset of these cases where one of the two components is real and the other is in a common fixed direction in the imaginary three-space, this is trivial, since then both 
$ \chi $ and 
$ \psi $ lie in the same complex plane, all the quantities in equations (\ref{chi.psi.relate}), (\ref{integrability.simpler.chi}), and (\ref{integrability.simpler.psi}) commute, and the obvious general solution to equation (\ref{chi.psi.relate}) is 
$ \chi = \psi $.

If such a two dimensional solution does not lie in a complex plane, then it lies in a plane in the totally imaginary three-space, and the axes of that space can be chosen such that
\begin{equation}
\chi = Q_{2} \chi_{2} + Q_{3} \chi_{3} = 
Q_{3} \left ( \chi_{3} + Q_{1} \chi_{2} \right ) = 
\left ( \chi_{3} - Q_{1} \chi_{2} \right ) Q_{3}
\label{imag.two.dim.chi}
\end{equation}
and
\begin{equation}
\psi = Q_{2} \psi_{2} + Q_{3} \psi_{3} = 
Q_{3} \left ( \psi_{3} + Q_{1} \psi_{2} \right ) = 
\left ( \psi_{3} - Q_{1} \psi_{2} \right ) Q_{3}
\label{imag.two.dim.psi}
\end{equation}
Since these will have the same amplitude, 
$ \chi_{2}^{2} + \chi_{3}^{2} = \psi_{2}^{2} + \psi_{3}^{2} $ and equation (\ref{chi.psi.relate}) becomes
{\samepage 
\begin{eqnarray}
\chi_{, \mu} \chi^{\dagger} & = & 
\psi^{\dagger} \psi_{, \mu} 
\nonumber \\
& = & \left ( \chi_{3} - Q_{1} \chi_{2} \right )_{, \mu} 
Q_{3} \left ( -Q_{3} \right ) \left ( \chi_{3} + 
Q_{1} \chi_{2} \right ) 
\nonumber \\
& = & \left ( \psi_{3} - Q_{1} \psi_{2} \right ) 
\left ( -Q_{3} \right ) Q_{3} \left ( \psi_{3} + 
Q_{1} \psi_{2} \right )_{, \mu} 
\nonumber \\
& = & \left ( \chi_{3} - Q_{1} \chi_{2} \right )_{, \mu} 
\left ( \chi_{3} + Q_{1} \chi_{2} \right ) 
\nonumber \\
& = & \left ( \psi_{3} - Q_{1} \psi_{2} \right ) 
\left ( \psi_{3} + Q_{1} \psi_{2} \right )_{, \mu} 
\nonumber \\
& = & \left ( \psi_{3} + Q_{1} \psi_{2} \right )_{, \mu} 
\left ( \psi_{3} - Q_{1} \psi_{2} \right ) 
\label{imag.two.dim.chi.psi.relate}
\end{eqnarray}}%
where the last step follows from the fact that all the quantities on the last three lines commute with each other. By inspection, the general solution to this equation is 
$ \chi_{2} = - \psi_{2} $, and 
$ \chi_{3} = \psi_{3} $. The integrability conditions are then also shown to be true just as easily as this solution was found.

Thus, the totally imaginary two dimensional case can be viewed as the case where 
$ \chi $ and 
$ \psi $ lie in the same two dimensional plane in the imaginary three-space, and they are reflections of each other through an arbitrary plane containing the origin and the normal vector at the origin to the plane containing 
$ \chi $ and 
$ \psi $. This last generalization follows since the original choice of space axes was made arbitrarily to simplify the form of 
$ \chi $ and 
$ \psi $, and to make the plane of reflection contain both the 
$ Q_{2} $ and 
$ Q_{1} $ axes as well. That specific case can be rotated back into a general plane in the imaginary three-space. It should be noted that all the two dimensional cases for 
$ \chi $ and 
$ \psi $ produce linear relations between them, along with the two linear Klein-Gordon forms\footnote{These Klein-Gordon forms may become nonlinear in strong gravitational fields with significant 
$ \hat R $ magnitudes, something that also happens in standard Klein-Gordon field theory with the added conformal term\cite{rankin.9404023}.} of equations (\ref{quat.free.kg.chi}) and (\ref{quat.free.kg.psi}). Thus, the principle of superposition applies to these solutions, allowing wave packets to be formed by conventional superposition of plane waves\cite{schiff,rankin.1101.3606}.

Now after setting 
$ C = - 1 $, use equation (\ref{def.identity.q.b0}) to convert back to standard lab coordinates with dimensions using 
$ b_{0} = 6 [ ( m_{0}^{2} c^{2} ) / \hbar^{2} ] $ for some reference ``rest mass'' 
$ m_{0} $, along with the speed of light 
$ c $ and the reduced Planck's constant 
$ \hbar $. Then typical plane wave solutions in the imaginary three-space might be 
$ \psi = Q_{3} [ A e^{ Q_{1} ( k x - \omega t ) } ] $ with 
$ \chi = Q_{3} [ A e^{ - Q_{1} ( k x - \omega t ) } ] $, where 
$ A $, $ k $, and $ \omega $ are all constants, and
\begin{equation}
k^{2} = \frac{\omega^{2}}{c^{2}} - 
\frac{m_{0}^{2} c^{2}}{\hbar^{2}}
\label{kg.sol}
\end{equation}
The fact that 
$ ( b_{0} / 6 ) C = - [ ( m_{0}^{2} c^{2} ) / \hbar^{2} ] $ has been used with both equations (\ref{quat.free.kg.chi}) and (\ref{quat.free.kg.psi}) here.

All these two dimensional forms will now be shown to be adequate to include the set of standard Dirac ``free particle'' solutions\cite{rankin.1101.3606} for spin 
$ 1 / 2 $, so they are potentially {\it not} just of academic interest.

\subsection{Spinor - Quaternion Equivalences}
\label{spin.to.quat}

A two row complex spinor 
$ \psi $ has four real components, 
$ \psi_{0 R} $, 
$ \psi_{0 I} $, 
$ \psi_{1 R} $, and 
$ \psi_{1 I} $, where the first row of the spinor is 
$ \psi_{0 R} + \imath \psi_{0 I} $, and the second row is 
$ \psi_{1 R} + \imath \psi_{1 I} $. A real quaternion also has four real components, so define the {\it quaternion equivalent} version of 
$ \psi $ as
\begin{equation}
\psi = \psi_{0 R} Q_{0} - \psi_{1 I} Q_{1} + 
\psi_{1 R} Q_{2} - \psi_{0 I} Q_{3}
\label{def.qpsi}
\end{equation}
In matrix form, this quaternion becomes
\begin{equation}
\psi =
\left (
\begin{array}{lr}
\psi_{0 R} + \imath \psi_{0 I} & -\psi_{1 R} + \imath \psi_{1 I} \\
\psi_{1 R} + \imath \psi_{1 I} & \psi_{0 R} - \imath \psi_{0 I} 
\end{array}
\right )
\label{def.m.qpsi}
\end{equation}
The first column of this is simply the original spinor, and the second column has the standard conjugate relationships to the first column common to all quaternions\cite{rankin.1101.3606}, so this is indeed a quaternion equivalent to the original spinor. In both cases, 
$ \psi^{\dagger} \psi $ has the same value. If there are two different original spinors 
$ \zeta $ and 
$ \psi $, the translation of the {\it spinor inner product} 
$ \zeta^{\dagger} \psi $ into quaternions will {\it not} be simply 
$ \zeta^{\dagger} \psi $, but rather the translation becomes
\begin{equation}
\zeta^{\dagger} \psi \rightarrow 
\frac{1}{2} \left ( \, \zeta^{\dagger} \psi - Q_{3} 
\zeta^{\dagger} \psi Q_{3} \right )
\label{def.general.inner.prod}
\end{equation}
as may be verified by direct calculation. This is the complex projection of a quaternionic value into the 
$ Q_{0} $, $ Q_{3} $ complex plane\cite{rotelli}. Thus, the translated inner product of two different spinor wavefunctions is still the same complex number in the quaternion formulation that it is in spinors. This actually becomes {\it the definition of the spinor equivalent inner product} of two such quaternions. Furthermore, when 
$ \zeta = \psi $, this reduces to 
$ \psi^{\dagger} \psi $, as it must. It should be noted that this quantity (which might well be called the ``spinner'' product) differs from the standard inner product of two quaternions, which is merely the dot product of the two quaternion space four component vectors that each original quaternion represents. That standard inner product can be expressed as 
$ ( 1 / 2 ) ( \zeta^{\dagger} \psi + \psi^{\dagger} \zeta ) $, and it is actually the real part of the spinor equivalent inner product.

One additional feature is necessary to convert many spinor equations into their quaternion equivalent equations. Spinors are frequently multiplied by 
$ \imath = \sqrt{ - 1 } $ in spinor equations, which indicates that each of the two spinor rows is multiplied by 
$ \imath $. But to multiply the same spinor rows by 
$ \imath $ in a quaternion equation, it is also necessary to multiply the two rows of the second quaternion column by 
$ - \imath $ as part of the same operation. But right multiplying any product by 
$ - Q_{3} $ produces exactly that result. The notation adopted to express this is that in the translation from spinor to quaternion equations, 
\begin{equation}
\imath \rightarrow - \left ( | Q_{3} \right )
\label{trans.i.to.quat}
\end{equation}
The operator 
``$ | $'' is called the barred operator, and it indicates that the quantity associated with it (the 
$ Q_{3} $) multiplies the product containing it from the {\it far right}\cite{rotelli,schwartz,rankin.1101.3606}. Thus, it produces the quaternion equivalent of multiplying the original spinor expression by 
$ \imath $. In that regard, note also that 
$ [ - ( | Q_{3} ) \psi ]^{\dagger}  = [ - \psi Q_{3} ]^{\dagger} = 
Q_{3} \psi^{\dagger} $, so that the 
$ Q_{3} $ must multiply the quaternion conjugate result from the left in order to maintain the correct equivalence to spinor results after multiplication. This should be expected, and can be facilitated further if necessary by defining a reversed barred operator\cite{rankin.1101.3606} that specifies multiplication from the far left in a product. That reversed barred operation is specified by 
$ ( Q_{3} \| ) $ when necessary, in obvious analogy to 
$ - ( | Q_{3} ) $.

\subsection{``Free Particle'' Dirac Solutions and Spin}
\label{free.dirac}

The standard solutions for the Dirac free particle are well known in a standard Dirac representation\cite{drell}. Those solutions may easily be reexpressed in a chiral representation in which the Dirac four row spinor is separated into a pair of two row spinors\cite{rankin.1101.3606}, either of which alone can describe the system by using the second order form of Dirac's equation, and using the appropriate first order Dirac equation to define the other two row spinor\cite{sakurai} if it is needed. If the upper two rows are chosen as the basic spinor, the standard free particle forms become
{\samepage 
\begin{eqnarray}
\xi_{u +} & = & 
\frac{1}{\sqrt{2}} \, 
e^{ - \imath \omega t } \left [
\begin{array}{r}
1 \\
0
\end{array}
\right ] , \, 
\xi_{d +} = 
\frac{1}{\sqrt{2}} \, 
e^{ - \imath \omega t } \left [
\begin{array}{r}
0 \\
1
\end{array}
\right ] , 
\nonumber \\
\xi_{u -} & = & 
\frac{1}{\sqrt{2}} \, 
e^{ \imath \omega t } \left [
\begin{array}{r}
1 \\
0
\end{array}
\right ] , \, 
\xi_{d -} = 
\frac{1}{\sqrt{2}} \, 
e^{ \imath \omega t } \left [
\begin{array}{r}
0 \\
1
\end{array}
\right ]
\label{bd.free.rest.chiral.eta}
\end{eqnarray}}%
The corresponding quaternionic forms of 
$ \xi $ are
{\samepage 
\begin{eqnarray}
\xi_{u +} & = & 
\frac{1}{\sqrt{2}} \, 
e^{ Q_{3} \omega t } , \, 
\xi_{d +} = 
\frac{1}{\sqrt{2}} \, 
Q_{2} e^{ Q_{3} \omega t } , 
\nonumber \\
\xi_{u -} & = & 
\frac{1}{\sqrt{2}} \, 
e^{ - Q_{3} \omega t } , \, 
\xi_{d -} = 
\frac{1}{\sqrt{2}} \, 
Q_{2} e^{ - Q_{3} \omega t }
\label{bd.free.rest.chiral.eta.quat}
\end{eqnarray}}%
where in both sets of equations, the subscripts 
``$ u $'' and 
``$ d $'' refer to spin up or down, and the 
``$ + $'' and 
``$ - $'' refer to positive or negative energy. Clearly the quaternionic forms of this set all satisfy equation (\ref{quat.free.kg.psi}) (with 
$ \psi = \xi \, $) in the flat space ``free particle'' case with 
$ k = 0 $, and 
$ \omega $ then given by equation (\ref{kg.sol}). They consist of two dimensional solutions in every case, and thus each has a corresponding 
$ \chi $ given by equation (\ref{chi.psi.relate}). Therefore they form a valid set of solutions to the constraint of equation (\ref{def.identity.q}). It should also be noted that this entire set can be generalized to represent free particle motion rather than just the case of the free particle at rest. A substitution such as 
$ \omega t \rightarrow \omega t - k z $ in all cases in the set generalizes it to a moving particle.

Now in standard Dirac theory, these solutions may be superposed to create new solutions such as superpositions of spin up and down. In quaternions, some of these combinations are still two dimensional with a known matching 
$ \chi $ from the rules already given, but some are fully four dimensional in quaternions, such as the superposition 
$ \xi = a \xi_{u +} + b \xi_{d +} $. However, this case can also easily be shown to have commuting right eigenvalues, which satisfies the integrability conditions. Indeed, the matching 
$ \chi $ for this case is easily shown to be the sum of the individual 
$ \chi $'s for each part. Other superpositions, such as 
$ a \xi_{u +} + b \xi_{d -} $ will have a 
$ \chi $ made up of the difference of the individual 
$ \chi $'s. In these ways, these linear superpositions are also solutions to the constraint equation (\ref{def.identity.q}) in the free space or ``free particle'' case. Thus, the constraint continues to show resemblances to particle mechanics just as it did with the original Weyl action in section \ref{discuss}. In this case, the similarity includes some properties of spin 
$ 1 / 2 $ quantum mechanics, and not just classical mechanics.

However, since only gauge invariant results have any real meaning, the above is still only a formal mathematical result rather than a direct statement about physics.\footnote{The wavefunction is {\it not} directly measurable, being part of the intrinsic standard of self measure (literally the ``gauge'') in the structure.} The simplest gauge invariant result that is available is the calculation of 
$ \hat v_{\mu} = \psi_{, \mu} \psi^{-1} = \xi_{, \mu} \xi^{-1} $. For a ``particle'' at rest 
($ k = 0 $), only 
$ \hat v_{0} \neq 0 $, and
{\samepage 
\begin{eqnarray}
\hat v_{u + \, 0} & = & Q_{3} \frac{\omega}{c} , \, 
\hat v_{d + \, 0} = - Q_{3} \frac{\omega}{c} , 
\nonumber \\
\hat v_{u - \, 0} & = & - Q_{3} \frac{\omega}{c} , \, 
\hat v_{d - \, 0} = Q_{3} \frac{\omega}{c}
\label{vhat.0.free}
\end{eqnarray}}%
Thus as far as these gauge invariant quantities are concerned, a flip in ``spin'' direction is equivalent to a flip in the sign of the ``energy''. Also note that the similar eigenvalues in the Dirac spinors are taken from 
$ \xi_{, \mu} = \lambda_{\mu} \xi $, where the 
$ \lambda_{\mu} $ are the eigenvalues. In the spinors, the eigenvalues do {\it not} flip sign under a flip in spin direction. Thus differences with Dirac formalism do appear. However, this particular difference comes directly from the translation from equation (\ref{bd.free.rest.chiral.eta}) to equation (\ref{bd.free.rest.chiral.eta.quat}), since 
$ Q_{3} $ does not commute with 
$ Q_{2} $ , but rather, anticommutes. The sign reversal vanishes if right eigenvalues 
$ \alpha_{\mu} $ are used in the quaternion case in the equation 
$ \xi_{, \mu} = \xi \alpha_{\mu} $ instead of left eigenvalues, but the equation 
$ \hat v_{\mu} = \xi_{, \mu} \xi^{-1} $ selects the left eigenvalue, not the right. Nevertheless, it will now be shown that superpositions of the 
$ \xi $ spin up and spin down cases (for the same energy), tell a much more detailed story than these deceptively few examples have, and will produce a 
$ \hat v_{0} $ that behaves much as spin 
$ 1 / 2 $ might be expected to behave.

Let
\begin{equation}
\xi = \cos{ ( \theta / 2 ) } \, \xi_{u +} + 
\sin{ ( \theta / 2 ) } \, \xi_{d +} e^{- Q_{3} \phi }
\label{superpose.rotated}
\end{equation}
where 
$ \theta $ is the spherical coordinate polar angle measured from the positive 
$ Q_{3} $ axis, and 
$ \phi $ is the azimuthal angle measured counterclockwise from the positive 
$ Q_{1} $ axis toward the positive 
$ Q_{2} $ axis, looking down from the positive 
$ Q_{3} $ axis.\footnote{This is based on the treatment of spin 
$ 1 / 2 $ on page 410 of Margenau and Murphy\cite{m.and.m}, {\it after} it is corrected for consistency with the (easily verified) correct results of page 406 by letting 
$ \phi \rightarrow - \phi $ on page 410.} Then a straightforward calculation of 
$ \hat v_{0} = \xi_{, 0} \xi^{-1} $ gives
\begin{equation}
\hat v_{0} = \frac{\omega}{c} \left ( Q_{1} \sin{\theta} 
\cos{\phi} + Q_{2} \sin{\theta} \sin{\phi} + Q_{3} 
\cos{\theta} \right )
\label{vhat0.quat.direction}
\end{equation}
The zeroth component of 
$ \hat v_{\mu} $ actually sweeps out a continuous change in direction in the purely imaginary quaternions to whatever direction the spherical coordinate angles specify. That is also the spin up direction in regular three space if the eigenspinors of 
$ \sigma_{z} $ are transformed and superposed via the exact spinor analog of the transformation and superposition of 
$ \xi_{u +} $ and 
$ \xi_{d +} $ in equation (\ref{superpose.rotated}) above\cite{m.and.m}. Thus, the quaternion imaginary three space direction of 
$ \hat v_{0} $ exactly tracks the spin up direction in ordinary three space. This is evidently a direct way to visualize spin direction via the quaternion space direction of this eigenvalue. In contrast, the 
($ \phi = 0 $ case) superposition 
$ \xi = \cos{ ( \theta / 2 ) } \, \xi_{u +} + \sin{ ( \theta / 2 ) } \, 
\xi_{u -} $ fails even to produce any well defined eigenvalue {\it except} at the extreme values of 
$ \theta = 0 $ and 
$ \theta = \pi $, which are the cases without superposition.\footnote{This absence of a precise eigenvalue because of oscillating cross terms and even infinite values in 
$ \hat v_{\mu} = \xi_{, \mu} \xi^{-1} $, is the common case when there are superpositions inside 
$ \xi $. The unusual, well defined eigenvalues of the spin superpositions indicate that the spin superpositions still represent an eigenstate.} That shows that the spin and energy cases are actually quite different, in spite of the agreement in eigenvalues at the extreme polar angles. At the same time, the rest of the material above shows that clear, gauge invariant spin 
$ 1 / 2 $ behavior does emerge in this framework via these solutions to the constraint equation (\ref{def.identity.q}).

Thus the equivalence to the free particle Dirac equation solutions is {\it not} just a mathematical formality. The surprise is that the spin 
$ 1 / 2 $ behavior appears directly in the imaginary quaternion space direction of an eigenvalue or eigenvalues previously unassociated with spin direction in the spinor Dirac formalisms\cite{schiff}, but rather associated with energy-momentum. Here, that ``energy-momentum'' set of eigenvalues remains,\footnote{True energy-momentum in this structure requires collaboration between these eigenvalues, and the appropriate terms in the stress tensor. A full example set is mapped out elsewhere for the case restricted to the complex plane\cite{rankin.9404023} with 
$ C = - 1 $. When that example set is restricted to the purely positive energy density field, the eigenvalues and stress tensor correlations match standard Klein-Gordon theory. For 
$ \hat \beta = 0 $, the ``vacuum'' becomes more of a ``virtual particle'' which can possess ``spin,'' but not real energy.} but is directly combined with the spin. Thus, the resemblance between the constraint and particle mechanics clearly continues, and is even enhanced with the quaternionic action of this section.

\section{Self Measurement and Mechanics}
\label{measure.mech}

The gauge invariant variables of this model embody the self measuring properties of Weyl-like geometries with scalar curvature 
$ B \neq 0 $. The scalar curvature becomes an intrinsic ruler by which the structure creates gauge invariant quantities suitable for physics. The geometry is thus literally, ``Self gauging,'' and displays that property through the intrinsic constraint of equation (\ref{def.identity.q}) between the gauge invariant variables. That constraint very closely resembles known equations from particle mechanics in combined gravitational and gauge fields, including classical mechanics, and even spin 
$ 1 / 2 $ quantum mechanics. Does this suggest a path to unify quantum mechanics, gauge fields, and General Relativity, all in one, unified geometric structure?

Objections to this suggestion are almost too obvious. First and foremost, there are the exotic, negative energy density field configurations allowed so far in this structure. Solving those requires either a sensible action principle that tames those exotic energy densities, or equivalently, discovery of a constructive and stable role for them that matches observed physics.\footnote{The Newtonian gravitational field is a negative energy density field that successfully matches observed physics in its realm of applicability.} Neither has yet been demonstrated that I know of.

The second obvious problem is that the similarities to mechanics are for single particle mechanics. Current physics is built upon quantum field theory which is a many body theory\cite{guidry}. The only theory of many body relativistic quantum mechanics that I know of that might fit naturally into this geometry is the, ``Many-amplitudes,'' theory of relativistic quantum mechanics explored by Egon Marx\cite{egon}. That might suggest a path to overcome this objection.

Third, particles come with many different rest masses, and the second order Dirac equation has a spin interaction term missing from the current structure\cite{rankin.1101.3606}. These tie together because at least in the complex plane, or for quaternionic cases projected into the complex plane, the introduction of 
$ \hat a_{\mu \nu} $, a self dual, antisymmetric component to the metric tensor may address both problems at once\cite{rankin.caqg,rankin.1101.3606}. The full metric tensor becomes 
$ \hat m_{\mu \nu} = \hat g_{\mu \nu} + \hat a_{\mu \nu} $, while its inverse 
$ \hat M^{\mu \nu} $ is defined by 
$ \hat M^{\mu \lambda} \hat m_{\lambda \nu} = \delta^{\mu}_{\nu} $. These give 
$ \hat M^{\mu \nu} = ( \hat g^{\mu \nu} - \hat a^{\mu \nu} ) / [ 1 + ( \hat a / 4 ) ] $, where all indices are raised and lowered using 
$ \hat g_{\mu \nu} $, and 
$ \hat a = \hat a^{\mu \nu} \hat a_{\mu \nu} $. In the resulting generalization of the constraint of equation (\ref{def.identity.q}), a spin interaction term is added, and the quantity 
$ C \rightarrow C \, [ 1 + ( \hat a / 4 ) ] $. But this allows the scalar plus pseudoscalar quantity 
$ \hat a $ to control the ``rest mass'', and to play much the same role as a Higgs Field in standard quantum field theory\cite{guidry}. Indeed, the zero order approximation is 
$ \hat a = - K^{2} $ where 
$ K $ is a real constant, and this is also the case that produces the correct Dirac-like spin interaction term. To fit that term, the value 
$ K = 4 $ seems to be the appropriate base Dirac case, as detailed in a separate paper\cite{rankin.1101.3606}. In the same framework, 
$ C = 1 $, since 
$ 1 + ( \hat a / 4 ) = - 3 $. This zero order approximation for 
$ \hat a $ would generalize to higher orders via 
$ \hat a = - \hat \lambda^{2} $, where 
$ \hat \lambda $ is a real variable field,\footnote{The antisymmetric part to the metric may not have a large amount of freedom if the structure is to be consistent in its self measurement. This is detailed in a separate paper\cite{rankin.1101.3606}.} one that obviously will now affect the ``rest mass,'' allowing for more than a single ``rest mass'' value.\footnote{Also, for 
$ \hat \lambda^{2} < 4 $, inflationary cosmologies are possible\cite{taylor.rankin}.}

However, whatever objections like these are raised, it's also still true that multiple coincidences in form and content appear between the internal constraint of these self measuring Weyl-like geometries, and known equations of classical and quantum particle mechanics in the gravitational and gauge fields of this structure. This includes close parallels in the form of the stress tensors and four currents of the gravitational and gauge fields\cite{rankin.caqg,rankin.9404023,rankin.1101.3606}. Furthermore, these familiar wave mechanical equation forms are known to produce sharp laboratory spectral lines that can used in lab measurements\cite{schiff}. Thus an abstract theory of a self measuring geometry points directly to a recognizable laboratory phenomenon, spectral lines which can be used in actual lab measurements. This is unlikely to be just coincidence. The scale of some predicted general relativistic effects will now be seen to add to the novelties.

\section{Scale of Curvature Related Effects}
\label{curvature}

As long as general relativistic effects involve only the (dimensionless) metric tensor, this model introduces no new scale for those effects. However, general relativistic quantities can also involve a nonzero curvature tensor 
$ \hat R^{\gamma}_{\mu \tau \sigma} $. Since that includes derivatives with respect to the coordinates, the dimensionless quantity used in this model is not quite the same as the usual expressions for curvature which still have units. Thus, this model will allow simple, new estimates to be made of scales at which this dimensionless curvature tensor cannot be ignored, and since it is dimensionless, those scales may have physical significance. But, it's also true that some important equations involving physical effects of the Riemann tensor will {\it not} affected by this scale factor dependent, dimensionless result. The scale factor completely cancels from the equation of geodesic deviation\cite{wald} when it is factored out to reexpress quantities in laboratory units, so geodesic deviation is clearly not affected. What follows will attempt to clarify these different possibilities.

\subsection{Ricci Tensor}
\label{ricci}

Einstein's equation (\ref{quat.einstein.result}) gives the Ricci tensor via the Einstein tensor, which is determined by the stress-energy tensor. The traditional Reissner-Nordstr\"{o}m solution for a point electronic charge\footnote{The problem of confining an actual charge in a volume of very small radius\cite{guidry} is not addressed here.} of magnitude 
$ e $, easily fits into the complex plane in this model\cite{rankin.9404023,rankin.1101.3606} (with the real part 
$ \hat y_{0 \mu \nu} = 0 $), and gives immediate estimates for the scale of 
$ \hat R_{\mu}^{\nu} $. It is easily shown that the diagonal components of 
$ \hat R_{\mu}^{\nu} $ are of unit magnitude at a radius from the central charge (in laboratory, Gaussian units)\footnote{Einstein's equations for the Reissner-Nordstr\"{o}m solution give a unit magnitude for diagonal components of 
$ \hat R_{\mu}^{\nu} $ when 
$ j^{2} ( g^{2} / 4 ) [ e^{2} / ( 2 b_{0}^{2} r_{L}^{4} ) ] = 1 $, where 
$ j^{2} $ is given by equation (\ref{def.j2}). Notice that 
$ g^{2} $ eventually cancels from this condition.}
\begin{equation}
r_{L} = \left ( \frac{e^{2}}{\hbar c} \right )^{1 / 4} 
\sqrt{ {\cal L}_{P} \, \frac{1}{\sqrt{b_{0}}} }
\label{rlab}
\end{equation}
In this, 
$ {\cal L}_{P} = \sqrt{ ( \hbar G ) / c^{3} } $, the well-known Planck length. Thus, the Ricci tensor is significant at a radius which is roughly the geometric mean of the Planck length, and the inverse of the square root of the scale factor 
$ b_{0} $.

This shows that the scale at which the Ricci tensor is significant depends explicitly and rather strongly on the choice made for the scale factor. If the scale factor is Planck scale, then the Ricci tensor becomes significant at a Planck scale radius, which is also roughly the radius at which the metric tensor begins to significantly deviate from flat space\cite{rankin.ijtp}. {\it However}, if the scale factor is atomic scale, then the Ricci tensor becomes significant at a {\it much} larger radius, a perhaps unexpected result. If 
$ b_{0} = ( 2 m_{0}^{2} c^{2} ) / \hbar^{2} $, and 
$ m_{0} $ is set to the electron mass (which would be consistent with the similarities to mechanics noted in section \ref{measure.mech} when 
$ C = 1 $ and 
$ \hat a = - 16 $), then 
$ r_{L} \approx 0.6 \cdot 10^{- 22} cm. $ This radius sweeps out a circle of cross sectional area roughly 
$ 10^{- 44} cm.^{2} $ around the central charge as viewed from any direction. This is approximately the lower bound of observed neutrino interaction cross sections\cite{texas92}, again perhaps unexpected. The independence of the scale factor from the Planck length produces this separate, intermediate scale constructed from both.

However, as noted above, some important equations involving physical effects of the Riemann tensor are {\it not} affected by this scale factor dependent result. The most immediate example of an equation that {\it does} reveal effects would be the covariant divergence of the stress tensor, specifically the terms 
\begin{equation}
\left [ \hat \beta_{0 \| \gamma}^{\: \: \: \: \; \| \gamma} 
\delta_{\mu}^{\nu} - \hat \beta_{0 \| \mu}^{\: \: \: \: \; \| \nu} 
\, \right ]_{\| \nu} = 
\hat R_{\mu}^{\nu} \hat \beta_{0 , \nu}
\label{div.extra}
\end{equation}
The coupling of those (force density) terms to the covariant divergence of the rest of the stress tensor appears very dependent upon the magnitude of the Ricci tensor. They are decoupled when it vanishes, even though 
$ \hat \beta $ is the same quantity that appears elsewhere in the stress tensor.\footnote{However, any additional 
$ \hat \alpha \hat R $ term in the action density for any other scalar field 
$ \hat \alpha $, generates similar terms and coupling in the stress tensor and its covariant divergence.} Only the dimensionless quantity 
$ \hat R_{\mu}^{\nu} $ clearly quantifies this general relativistic variation in coupling.

\subsection{Scalar Curvature}
\label{scalarcurv}

The scale at which 
$ \hat R $ becomes significant can be estimated through the portion of the stress-tensor in equation (\ref{quat.einstein.result}) generated by the constraint terms. This estimate appears in another paper\cite{rankin.1101.3606} assuming a case limited to the complex plane. Positive energy density is assumed, and it is assumed that the wavefunction for the designated ``rest mass'' can be contained in a small enough (spherical) volume to reach the peak magnitude necessary to produce a unit magnitude 
$ \hat R $ inside that volume. The calculated radius of such a volume is within a few orders of magnitude of the radius above for significant 
$ \hat R_{\mu}^{\nu} $, and is smaller, at least when the electron rest mass is the mass involved throughout. The most obvious effect a significant scalar curvature has is through the 
$ \hat R $ term in the constraint\cite{rankin.9404023}, which plays the same role as the added conformal term in a Klein-Gordon equation\cite{wald}. That term is otherwise ignored.

\section{Summary}
\label{summary}

The well known Weyl geometry can be successfully extended to quaternionic gauge transformations and Yang-Mills fields provided half the nonmetricity is shifted into torsion. This is the primary finding of this paper. For nonvanishing scalar curvature, the geometry provides intrinsic gauge invariant variables which can correspond to measured quantities in physics, and which facilitate the extension to the quaternions by minimizing the number of noncommuting quantities in the equations. This is the second main topic of this paper.

Those gauge invariant variables are rendered dimensionless by expressing all coordinates as dimensionless quantities by use of a scale factor 
$ b_{0} $ with dimensions of inverse length squared. The kinematics of the geometry require the intrinsic gauge invariant variables always to obey a constraint. When examined together with some particular choices of action principle, that constraint shares many properties with well known equations from classical and quantum particle mechanics, including mechanics of a spin 
$ 1 / 2 $ Dirac particle. At the same time, those resemblances give the gauge invariant form of the torsion (and nonmetricity) the character of ``momentum-energy'' demanded of torsion in the literature\cite{gronwald.hehl}. These are additional main points of this paper.

Besides the constraint, the full equations of motion are developed given an action principle with a fully quaternionic Yang-Mills field. The action used is basically an Einstein/Yang-Mills action with some optional additional terms suggested by Weyl's original action. Cases limited to real variables, or the complex plane, are included. The constraint is necessarily included in the action, and generates field sources for both the Einstein and Yang-Mills fields. The resulting equations for an isolated electronic charge and mass suggest that unlike the metric tensor deviations from flat space, the (newly dimensionless) Ricci tensor in General Relativity may become significant at much greater distances than the Planck length, depending on the value assigned to the scale factor. However, many general relativistic results such as geodesic deviation are unaffected by this scaling.

Overall, this model extends Weyl's work into the quaternions, and that facet should not be controversial. It then offers tantalizing hints of deep links between the internal structure of Weyl-like geometries, classical and quantum particle mechanics, spin 
$ 1 / 2 $, and the self-measuring property of this type of geometry. In the Weyl-like geometry, these are directly integrated with the Einstein and Yang-Mills fields that share the unified structure, plus a possible self dual antisymmetric part to the metric to provide both a spin interaction, and a controller of ``rest mass.'' Nevertheless, substantial obstacles remain to such an approach overall, such as the necessity for a ``many body'' version of the formalism. The framework also is too narrow to embrace the full 
$ SU(3) \times SU(2) \times U(1) $ structure of the current Standard Model\cite{guidry}, although it does contain the 
$ SU(2) $ and 
$ U(1) $ gauge fields as subsets of the quaternionic Yang-Mills field. The quaternionic Yang-Mills field is 
$ SU(2) \times SU(2) $ structure.

\begin{acknowledgments}

I would like to thank Daniel Galehouse, Jim Wheeler, and Egon Marx for discussions that contributed to this paper.

\end{acknowledgments}




\end{document}